\documentclass[acmtog]{acmart}

\setcitestyle{square}

\usepackage{amsmath,amsthm,amssymb}
\usepackage{graphicx}
\usepackage{textcomp}
\usepackage{wrapfig}
\usepackage{subfig}
\usepackage{parskip}
\usepackage{color}
\usepackage{xspace}
\usepackage{overpic}
\usepackage{subfig}
\usepackage{wrapfig}
\usepackage{enumitem}
\usepackage{mathrsfs}
\usepackage{multirow}

\definecolor{turquoise}{cmyk}{0.65,0,0.1,0.1}
\definecolor{purple}{rgb}{0.65,0,0.65}
\definecolor{dark_green}{rgb}{0, 0.5, 0}
\definecolor{orange}{rgb}{0.8, 0.6, 0.2}
\definecolor{red}{rgb}{0.8, 0.2, 0.2}
\definecolor{brown}{rgb}{0.5, 0.16, 0.16}

\newcommand{\rz}[1]{{\color{black}#1}}

\newcommand{\revised}[1]{{\color{black}#1}}

\renewcommand{\textrightarrow}{$\rightarrow$}

\begin{document}
\title{P2P-NET: Bidirectional Point Displacement Net for Shape Transform}

\author{Kangxue Yin}
\affiliation{%
  \institution{Simon Fraser University}
}
\author{Hui Huang}
\affiliation{%
	\institution{Shenzhen University}
}
\author{Daniel Cohen-Or}
\affiliation{%
	\institution{Tel Aviv University}
}
\author{Hao Zhang}
\affiliation{%
	\institution{Simon Fraser University}
}

\renewcommand\shortauthors{K. Yin, H. Huang, D. Cohen-Or, and H. Zhang}

\begin{abstract}
We introduce P2P-NET, a {\em general-purpose\/} deep neural network which learns geometric
transformations between point-based shape representations from two domains, e.g., meso-skeletons and surfaces, 
partial and complete scans, etc. The architecture of the P2P-NET is that of a 
{\em bi-directional point displacement\/} network, which transforms a source point set to a prediction of the target point 
set with the same cardinality, and vice versa, by applying {\em point-wise\/} displacement vectors learned from 
data. P2P-NET is trained on {\em paired\/} shapes from the source and target domains, but 
{\em without\/} relying on point-to-point correspondences between the source and target point sets.
The training loss combines two uni-directional geometric losses, each enforcing a {\em shape-wise\/}
similarity between the predicted and the target point sets, and a cross-regularization term to 
encourage consistency between displacement vectors going in opposite directions.
We develop and present several different applications 
enabled by our general-purpose bidirectional P2P-NET to highlight the effectiveness, versatility, and potential 
of our network in solving a variety of point-based shape transformation problems.
\end{abstract}

%
%

\ccsdesc[500]{Computing methodologies~Computer graphics}
\ccsdesc[500]{Computing methodologies~Shape modeling}
\ccsdesc[500]{Computing methodologies~Shape analysis}

\setcopyright{acmcopyright}
\acmJournal{TOG}
\acmYear{2018}\acmVolume{37}\acmNumber{4}\acmArticle{152}\acmMonth{8} \acmDOI{10.1145/3197517.3201288}

\keywords{Point cloud processing, deep neural network, point-wise displacement, point set transform}





\begin{teaserfigure}
  \centering
  \includegraphics[width=\linewidth]{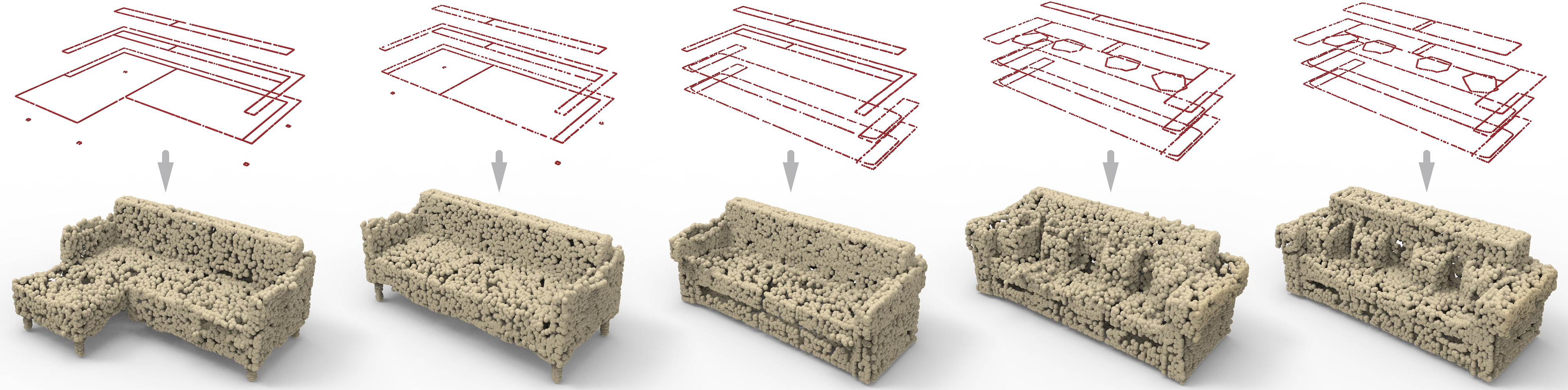}
  \caption{We develop a general-purpose deep neural network which learns geometric transformations between point sets, e.g., from cross-sectional profiles to 3D shapes, as shown. User can edit the  profiles to create an interpolating sequence (top). Our network transforms all of them into point-based 3D shapes. \newline}
  \label{fig:teaser}
\end{teaserfigure}

\maketitle




\section{Introduction}
\label{sec:intro}

\begin{figure*}[t!]
	\centering
	\includegraphics[width=0.9\linewidth]{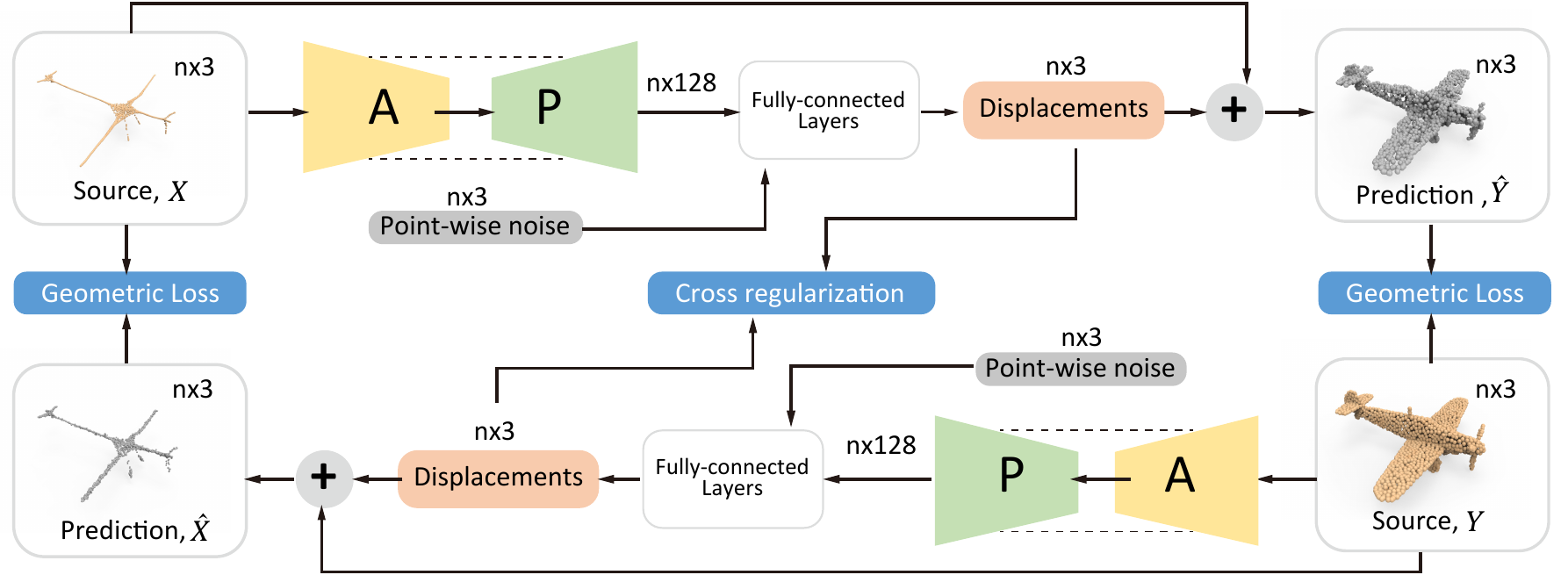}
	\caption{Network architecture of our bidirectional P2P-NET, \revised{which transforms a source point set to a prediction of the target point set with the same cardinality. Note that blocks A and P represent set abstraction layers and feature propagation layers, respectively, of PointNET++~\protect\cite{qi_nips2017}.}}
	\label{fig:archi}
\end{figure*}

Point primitives and point-based processing 
have attracted considerable interest from
the computer graphics community for many years~\cite{gross2007}. As a fundamental
shape representation, point sets are more compact than voxels and more flexible than
polygon meshes. They are immediately available as the default output from most 
3D shape acquisition devices. 
%
%
Recently, deep neural networks have been designed to learn global and multi-scale 
point set features for shape classification and segmentation~\cite{qi_cvpr2017,qi_nips2017}, as 
well as geometry (e.g., normal or curvature) estimation~\cite{guerrero2017pcpnet}. 
Image-driven generative models have also been trained to reconstruct point-based 3D object 
representations from single or multi-view images~\cite{fan2016point,lin_aaai2018,lun2017sketch}.

In this paper, we are interested in exploring how deep neural networks can benefit a new class of problems
in point-based graphics: {\em geometric transformations\/} between shapes represented by point sets. 
These shape transforms span a wide spectrum of applications. Some examples include 
transforms between shape skeletons and surfaces, incomplete and completed object scans, 2D contours 
and 3D shapes, and simplified and detailed surfaces. Our goal is to develop a {\em general-purpose\/} 
neural network that is capable of learning geometric transformations between point sets from two domains.
Recently, in computer vision, there has been a great deal of interest in solving a
similar problem for images, namely, designing general-purpose, end-to-end image-to-image 
translation networks~\cite{isola2017image,liu2017nips,zhu2017unpaired,yi2017dualgan}.


Most successes on generic image translation have been achieved for tasks leading to
``stylistic'' changes in images, without geometrically transforming the image content. These tasks
include translating between night and day images, artistic styles, material properties, etc. Under this 
setting, some recent works, such as CycleGAN~\cite{zhu2017unpaired} and DualGAN~\cite{yi2017dualgan} 
can train their translators {\em without\/} paired images, but they both rely on loss functions that 
measure pixel-to-pixel differences. Thus, while the images are not paired, the pixels are. However, when the translation
involves geometric changes, these methods have all encountered clear obstacles.

In our work, the distinction, as well as the key novelty, of the problem is that transforming
geometry {\em is\/} the goal. While pixel correspondences can be trivially defined between two images
depicting the same content (e.g., night vs.~day photos of the same location), pairing points which represent 
geometrically different shapes is, in general, far from straightforward. Even if the two sets of points were sampled 
from the same shape, e.g., a cross-sectional profile representation vs.~that of the whole (see Fig.~\ref{fig:teaser}), there may not always be a clear point-to-point correspondence 
between them. In \rz{our first attempt to develop a general-purpose transformation network for point sets\/}, and in contrast to CycleGAN and DualGAN, 
we rely on paired shapes in training, but do not require paired points.



Specifically, 
we design a {\em point-to-point displacement\/} network, coined {\em P2P-NET\/}, which transforms an input point set 
to an output set with the same cardinality, by applying {\em point-wise\/} displacement vectors learned from
data. The P2P-NET is trained under a weakly-supervised setting, where paired point sets which share some commonality
are provided, but not their point-wise mapping. In most of the applications considered, the commonality is that the
two point sets were sampled from the same shape. However, in many applications, there is no clear point-to-point 
correspondence between the two point sets, or the two sets could contain point samples acquired under different 
view settings or at different time instants.
Not requiring point-wise correspondences can significanly expand our capacity to collect training data for P2P-NET.

Given two domains of point set data, $\mathscr X$ and $\mathscr Y$, we introduce a {\em bidirectional\/} architecture to 
learn the two transformations $\mathscr X$-to-$\mathscr Y$ and $\mathscr Y$-to-$\mathscr X$, simultaneously, as shown
in Fig.~\ref{fig:archi}.
Given a source point set $X$, the (bidirectional) P2P-NET learns to predict a set of displacement vectors $\mathcal{I}_X$ that 
are applied to $X$ to obtain the predicted point set $\hat{Y}=X + \mathcal{I}_X$.
One objective of training the P2P-NET, defined by a {\em geometric loss\/}, is to make the prediction $\hat{Y}$ as close as possible to the
target {\em shape\/} represented by the point set $Y$. At the same time and along the opposite direction, the network also learns to
predict displacement vectors $\mathcal{I}_Y$, such that $Y + \mathcal{I}_Y$ is as close as possible to $X$, shape-wise.
In addition, we define a {\em cross regularization\/} loss which couples and
{\em mutually enhances\/} the two directional sub-networks, by encouraging {\em parallelism\/} between
the two sets of displacement vectors $\mathcal{I}_X$ and $\mathcal{I}_Y$. Our bidirectional P2P-NET is trained
with a combined loss consisting of two (directional) geometry terms and one cross regularization term. None of 
the loss terms requires a point-wise correspondence between $X$ and $Y$.

Along each direction of the P2P-NET, the network takes a set of 3D points as input and first learns a multi-scale feature 
per point through set abstraction and feature propagation layers of PointNet++~\cite{qi_nips2017}. \revised{The learned feature
vector for each point is then {\em concatenated\/} with an {\em independent\/} Gaussian noise vector. The set of
``noise-augmented'' feature vectors are fed into a set of fully connected layers, which produce a set of 
3D displacement vectors, one per input point, as shown in Fig.~\ref{fig:archi}.}

P2P-NET learns point set transforms {\em implicitly\/}. It does not learn how to directly displace each point in the source 
shape to its corresponding point in the target shape, since such point-to-point correspondences are not provided by training data. 
Instead, our network learns a mapping from point features (encoded as in PointNET++) to displacement vectors, which are applied 
to the source shape.
\revised{As a result, nearby points in the source shape, which often possess similar point features, tend to be mapped to
similar displacement vectors, which would impose a certain ``rigidity'' on the shape transforms that P2P-NET can learn.
The augmentation of independent per-point noise into P2P-NET alleviates this problem by providing added degrees of 
freedom to the displacements of individual points, allowing the network to learn a richer variety of transforms.}


The main contributions of our work include:
\begin{itemize}
\item The first general-purpose deep neural network designed to learn transformations between point-based shape representations.
\item Bidirectionality of the network with a novel cross regularization loss to mutually enhance two directional transforms.
\item Training losses defined without point-to-point correspondence, allowing the network to be trained under
weak supervision.
\end{itemize}

We demonstrate how noise augmentation, cross regularization, and bidirectionality improve the performance of 
our P2P-NET, as it carries out its learning tasks. We develop and present several different applications 
enabled by our general-purpose bidirectional P2P-NET to highlight the effectiveness, versatility, and potential 
of our network in solving a variety of point-based shape transformation problems. These include transforms
between skeletons and shapes, between skeletons and scans, between partial and complete scans, and finally
between cross-sectional profiles and 3D shapes.

\section{Related work}
\label{sec:related}

The literature on point-based graphics and the application of deep neural networks to solve graphics problems is vast. 
In this section, we only cover the most relevant works to P2P-NET.

\vspace{-5pt}

\paragraph{\bf{Point processing.}}
Point cloud, as a 3D shape representation, has shown its advantages and applications in geometry modeling~\cite{pauly2003shape}, rendering~\cite{alexa2003computing}, and computer animation~\cite{muller2004pointAnimation}. With the emergence of affordable 3D acquisition devices, point cloud data is widely captured, accumulated, and processed by many useful techniques, e.g., in point cloud filtering~\cite{mitra2003estimating}, resampling~\cite{huang2013edge}, and surface reconstruction~\cite{carr2001reconstruction,kazhdan2013screened,Berger2017}, among others.

\vspace{-5pt}

\paragraph{\bf{Point set displacement.}}
Several previous works have taken a displacement-based approach to process point sets. One such example is point
set skeletonization, where point samples gradually converge from a shape's surface to a skeletal structure.
The main challenge lies in how to deal with missing data
%
over the latent surface, e.g., when the point scan was acquired from a single view. Tagliasacchi et al.~\shortcite{tagliasacchi2009curve} propose a generalized rotational symmetry axis (ROSA) to extract curve skeletons from incomplete point clouds. Cao et al.~\shortcite{cao2010point} apply a Laplacian-based contraction to extract curve skeletons. Huang et al.~\shortcite{huang2013l1} introduce $L_1$-medial skeletons by adapting $L_1$-medians locally to an incomplete point set representing a 3D shape. Our P2P-NET leads to a 
data-driven approach to point cloud skeletonization and the bidirectional network is also trained for a novel 
task: {\em skeleton-to-shape\/} transform.

Another example is surface completion by evolving point samples to gradually fill missing data over the latent surface.
Representative methods include point cloud consolidation via LOP~\cite{lipman2007parameterization}
and WLOP~\cite{huang2009consolidation,preiner2014}. 
A more recent work that bridges point cloud skeletonization and consolidation~\cite{wu2015} spreads point samples regularly to cover 
gaps over surface regions via a joint optimization of surface and structure samples. Point cloud resampling can also be applied
for edge enhancements~\cite{huang2013edge}.
P2P-NET offers a data-driven approach to surface completion, via point displacements, which offers an alternative to
other learning-based surface completion techniques, such as the recent work by Dai et al.~\shortcite{dai2017complete}.

\vspace{-5pt}

\paragraph{\bf{Neural networks for point processing.}}
Neural networks excel in learning global features. A key development in connecting point sets to neural networks is PointNet~\cite{qi_cvpr2017}, which directly consumes unorganized point samples as input.
This is followed by PointNet++~\cite{qi_nips2017}, which enables hierarchical learning on point 
sets. In both cases, the input point set goes through point-wise or patch-wise feature transform followed
by feature aggregation, either globally or locally, so as to serve the tasks of shape classification 
and segmentation (i.e., patch classification).
%
Another multi-scale variant of PointNet, by Guerrero et al.~\shortcite{guerrero2017pcpnet}, is 
adapted for estimating local shape properties, such as normals and curvatures. Sung et 
al.~\shortcite{sung2017complementme} demonstrate the usefulness of PointNets for component 
suggestion in part-based shape assembly. 

Our P2P-NET is designed to solve a different class of problems, namely, point displacement based
shape transforms. While P2P-NET does employ the set abstraction and feature propagation layers of 
PointNet++~\cite{qi_nips2017} for feature learning, it combines the learned features 
with noise and trains a bidirectional network, with a novel loss term combining shape 
approximation and cross regularization, to obtain point displacement vectors. 

There have been several recent attempts at developing deep generative networks for point-based 
shapes. Fan et al.~\shortcite{fan2016point} design and train a neural network as a conditional sampler, 
which is capable of predicting multiple plausible 3D point clouds from a single input image. 
Multiple point clouds from different views have also been constructed as intermediate shape 
representations for the purpose of generating 3D shapes from 2D images and/or 
sketches~\cite{lun2017sketch,lin_aaai2018}.
Gadelha et al.~\shortcite{gadelha2017PCA} synthesize point-based 3D shapes in the space of shape coefficients, 
using a generative adversarial network (GAN). They build a KD-tree to spatially partition the points and then conduct PCA analysis to derive a linear shape basis and optimize the point ordering.
%

To the best of our knowledge, P2P-NET is the first deep neural network designed to learn 
geometric transformations between point-based shape representations. 

\vspace{-5pt}

\paragraph{\bf{Learning transformations.}} 
Several classical vision problems need to account for spatial transformations in images, e.g., 
recognizing objects undergoing deformations~\cite{jaderberg2015spatial} and 
synthesizing images under novel views~\cite{zhou2016view}, among others. Both
works are representative of applying deep neural networks for their respective tasks.
%
%
There has been considerable less effort on learning geometric transforms for 3D shapes. Recent attempts have
been made to learn to transfer surface details~\cite{berkiten2017detail}, decorative styles~\cite{hu2017}, and 
to predict piecewise rigid transformations of 3D objects~\cite{byravan2016se3}. In contrast, our work aims to develop
a {\em general-purpose\/} neural network for learning transformations between point-based 3D shapes.

\vspace{-5pt}

\paragraph{\bf{Paired vs.~unpaired training data.}} 
Analogous to our shape transform problem, is general-purpose image-to-image 
translation~\cite{isola2017image,liu2017nips,zhu2017unpaired,yi2017dualgan}, where point displacements can be
regarded as a counterpart to pixel-to-pixel transforms. An important feature of some of these recent 
works~\cite{liu2017nips,zhu2017unpaired,yi2017dualgan} is that the training does not require paired images.
On the other hand, these works, which rely on deep generative neural networks such as GANs, have only shown
success in color and texture transforms, e.g., for altering painting styles or material properties of imaged objects. 
Training these networks to deform objects geometrically in images has remained an unresolved challenge. In contrast,
P2P-NET is designed to learn geometric transforms between 3D shapes; 
it requires paired training data, but not paired points.

\vspace{-5pt}

\paragraph{\bf{Bidirectionality vs.~cycle consistency.}} 
Our design of the bidirectional P2P-NET drew inspirations from dual learning~\cite{yi2017dualgan} and 
the use of cycle consistency loss~\cite{zhu2017unpaired}. What is common about these works is that they all 
learn transforms between two domains. The cycle consistency loss is a clever way of dealing with the challenge 
of not having paired training data from the two domains. \rz{However, P2P-NET is not built on a cyclic loss.\/} 
With paired training data for P2P-NET, we can afford to define the two directional geometry losses, without the
inverse mappings. At the same time, the bi-directionality between the two transforms is taken advantage of, since 
we define the extra cross regularization loss to enhance the training.
Another distinction lies in how the loss functions are defined: the cycle consistency loss measures pixel-to-pixel
differences, while our geometry loss measures a {\em shape-wise\/} difference between two point sets.

\section{Bidirectional P2P-Net}
\label{sec:method}

\begin{figure*}[!t]
	\centering
	\includegraphics[width=0.9\linewidth]{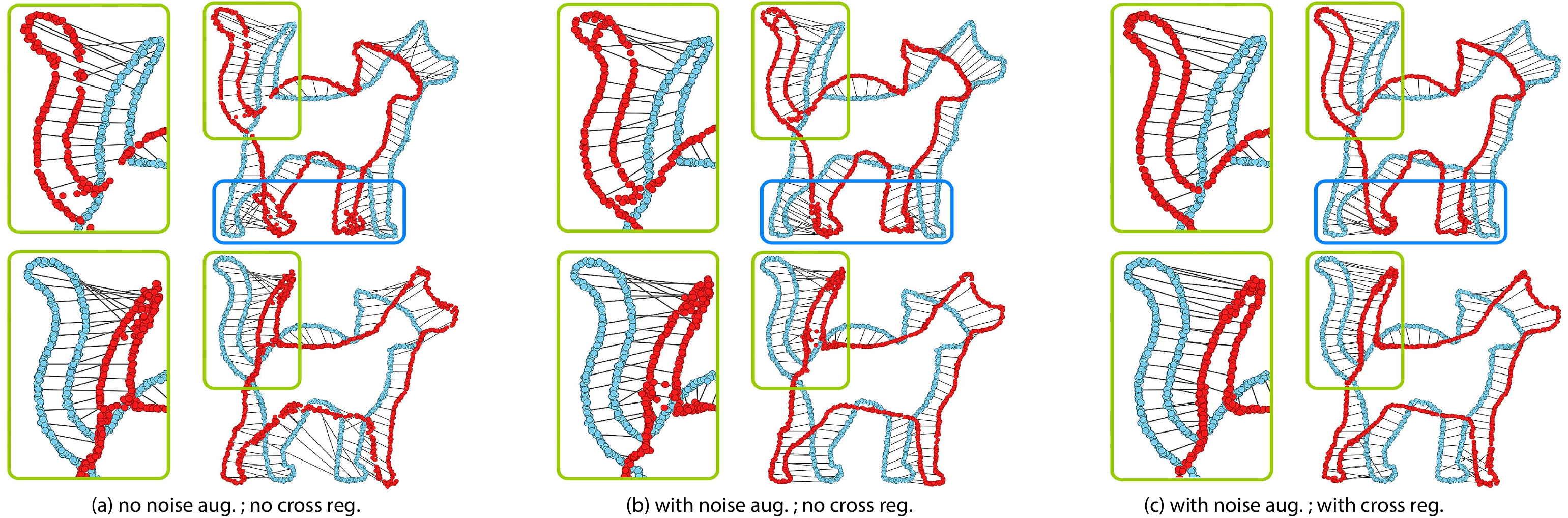}
	\caption{\revised{An ablation study on cat-dog transforms. P2P-NET was trained on a dataset synthesized by randomly rotating and scaling a pair of 2D point sets representing the shapes of a dog and a cat, respectively. Top row: dog to cat. Bottom row: cat to dog. The source shape is always shown in blue and prediction in red, and only 10\% of the displacements are displayed. Light green and blue boxes highlight areas with visible improvements.}}
	\label{fig:catdog}
\end{figure*}

The architecture of bidirectional P2P-NET is illustrated in Fig.~\ref{fig:archi}.
Our training set consists of paired point sets $\{X, Y\}$,  with prior relations among them. However, the transformations between two sets $X$ and $Y$ are latent and difficult to model explicitly.
For instance, $X$ can be a single-view point scan of a chair, and $Y$ contains complete surface samples of the same chair. As another example, shown in Fig.~\ref{fig:catdog}, a 2D 
point set sampled from a dog shape is transformed to represent the shape of a cat and vice versa.

To realize a bidirectional architecture on unordered point sets, we develop a geometric loss (Section~\ref{sec:matching}) that is order-invariant. Since $X$ and $Y$  are not in dense correspondences, i.e., point-wise, we need to further regularize the loss to balance the mapping and the global distribution of the displacements. To this end, the loss of bidirectional transformations is tightly coupled with a cross regularization (Section~\ref{sec:regularization}) that maximizes the parallelism between displacements from $X$-to-$Y$ and displacements from $Y$-to-$X$. 


P2P-NET consists of two network branches in two opposite directions. At each branch, the network first learns a multi-scale feature for each point, using layers of PointNet++~\cite{qi_nips2017}. The input point set is down-sampled and point features are constructed in multiple levels with set abstraction layers (marked with A in Fig.~\ref{fig:archi}). Point-wise multi-scale features are then produced with the feature propagation layers (marked with P in Fig.~\ref{fig:archi}).
Next, the multi-scale point-wise feature vectors are concatenated with the same number of noise vectors, one per point. Each noise vector is an independent Gaussian noise vector of length 32.  
Finally, the feature-noise vectors are fed to a set of fully connected layers that output displacement vectors $\mathcal{I}_X$. In the end, the network yields the predicted point set $\hat{Y}=X + \mathcal{I}_X$. 
See Appendix~\ref{sec:appendix}.1 for a more detailed description of the network architecture.

\begin{figure}[!t]
	\centering
	\includegraphics[width=0.95\linewidth]{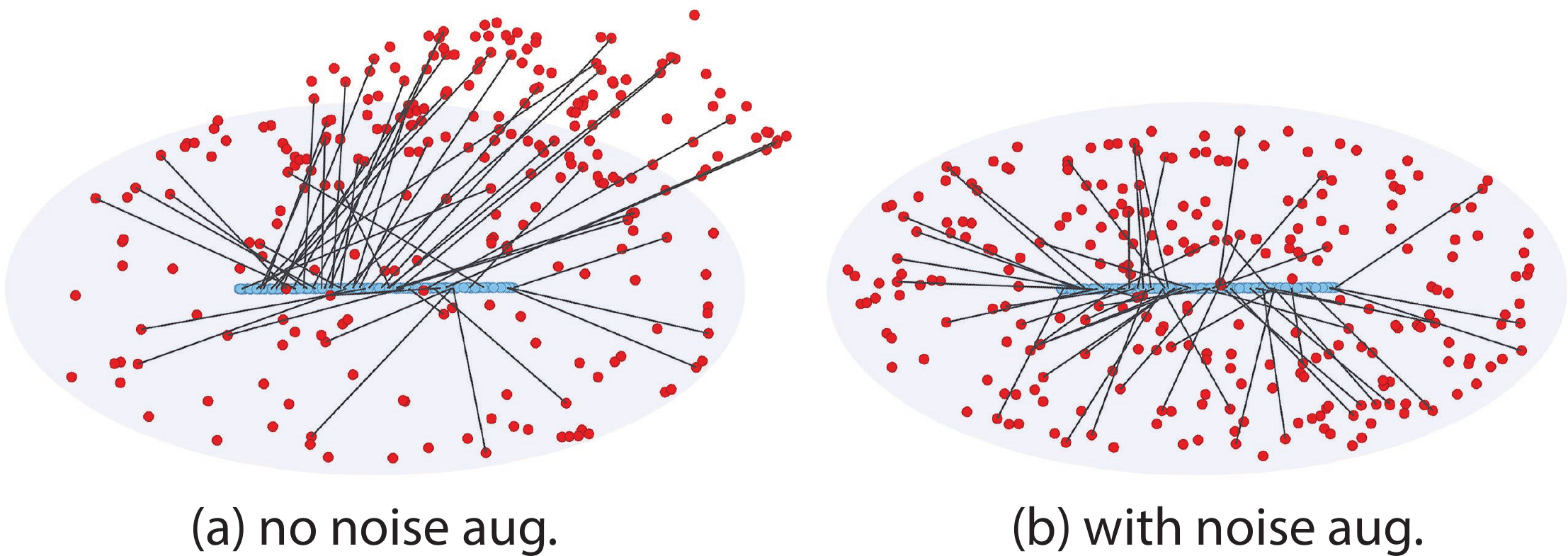}
	\caption{\revised{Noise augmentation allows P2P-NET to learn to transform points along a straight line (blue dots in the center) to points distributed over an elliptical disk (red dots) more effectively. For a clearer visualization, we only show 20\% of the displacement vectors (black lines) which are randomly chosen. Note also that the cross regularization term is not employed.}}
	\label{fig:noise_ab}
\end{figure}

\revised{
Noise augmentation in our P2P-NET adds new dimensions to the feature vectors. These newly added (noise) dimensions inject new degrees of freedom, when the network learns to map
point features to displacements during training. This effectively neutralizes an ``overfitting'' of the point displacements to point features and enables more variation in the displacements.
The appended noise vectors are independent to each other, allowing each point to train for its own variation to further improve the versatility of P2P-NET.
Furthermore, since the noise introduces stochasticity into the network, we can feed an input point set multiple times during testing, to obtain a dense output point set;
see Fig.~\ref{fig:crosssection} for an example.

Fig.~\ref{fig:noise_ab} illustrates the effect of noise augmentation using a toy example, where P2P-NET is trained to transform points along a straight line to points distributed over an 
elliptical disk. The network was trained on 1,000 line-disk pairs with random scales and orientations. Since P2P-NET only learns a mapping from point features to displacements, it 
intrinsically respects the smooth variation of the point features along the straight line. However, the transform task at hand sets a conflicting goal, which is enforced by the geometry
loss in the network: map smoothly varying point features to ``non-smooth'' displacement vectors, so that the output points can be well-distributed over a disk. Without noise augmentation, 
P2P-NET would respect the point features relatively more rigidly. As shown in Fig.~\ref{fig:noise_ab}(a),
the network struggles to fulfill the two conflicting goals and produces many similar displacement vectors, causing some points to overshoot over the disk boundary.
In contrast, with noise augmentation, the points have added degrees of freedom to be mapped to non-smooth displacement vectors to minimize the geometry loss.
The final result is a significantly better point distribution over the disk without overshooting, as shown in Fig.~\ref{fig:noise_ab}(b).

}

In addition to the geometric losses defined and enforced at the two input/output ends of the bidirectional P2P-NET, the aforementioned cross regularization over the point 
displacements strengthens the coupling between the two directional networks. The ablation study shown in Fig.~\ref{fig:catdog} demonstrates the impact of both
noise augmentation and cross regularization on a less toyish example: transforming between points representing the shapes of dogs and cats.

\begin{figure*}[!t]
	\centering
	\includegraphics[width=0.85\linewidth]{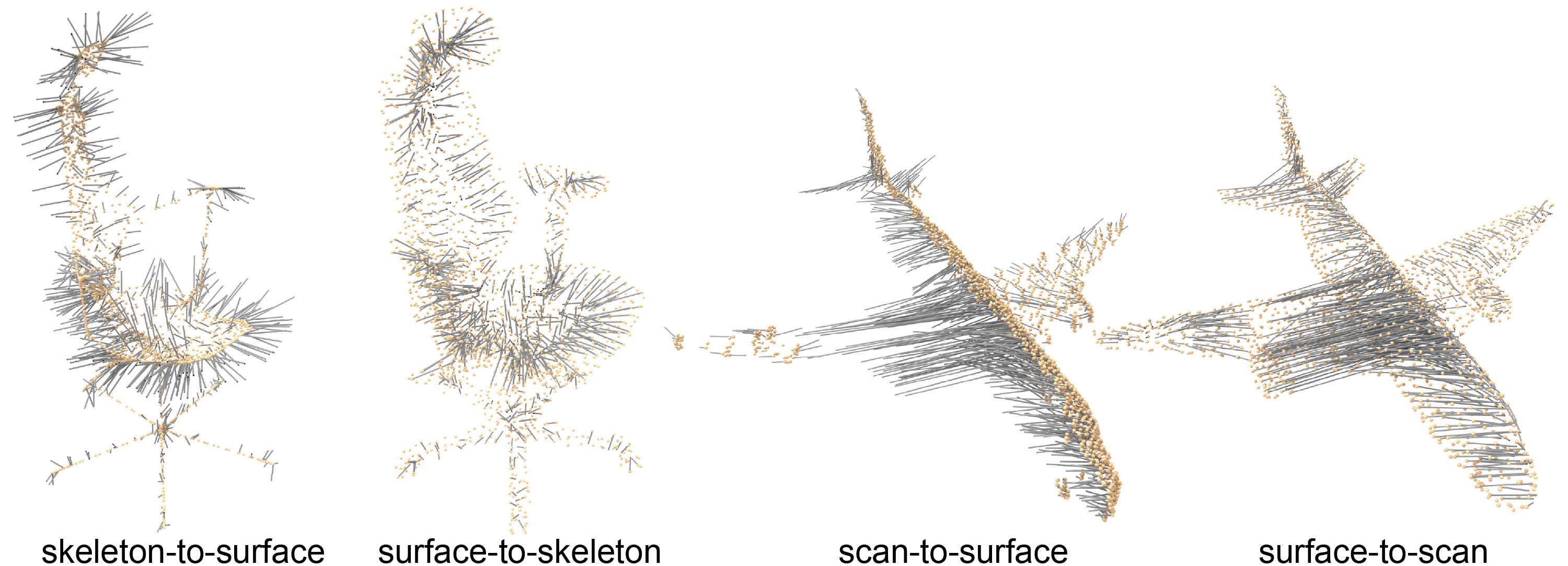}
	\caption{Visualization of vectors (grey lines) depicting point-wise displacements learned by P2P-NET for various domain mappings, where the source point sets are rendered in orange. Note that for ease of visualization, only 30\% of the vectors are displayed and we do not show the predicted target point sets.}
	\label{fig:displace_visul}
\end{figure*}

\subsection{Geometric Losses }
\label{sec:matching}

To measure the geometric difference between the predicted and target point sets,  the network is trained with a loss that consists of two terms. One term penalizes points that do not match with the target shape, and the other term measures the discrepancy of the local point density between two corresponding point sets.

The shape matching loss computes the sum of differences between the shape of transformed point set $\hat{Y}=X + \mathcal{I}_X$ and the shape of target point set $Y$, vice versa between $\hat{X}=Y + \mathcal{I}_Y$ and $X$, by searching the closest point from target point set for each displaced source point:
\begin{eqnarray*}
	L_\text{shape}(\hat{Y}, Y) = \sum\limits_{p \in Y} \underset{q\in \hat{Y}}{\min} \ d(p, q)  + \sum\limits_{q \in \hat{Y}} \underset{p\in Y}{\min} \ d(p,q) ,
	\label{eq:shape}
\end{eqnarray*}
where $d(p,q)$ measures $L2$ distance between points $p$ and $q$.

This symmetric shape matching term is close to the Hausdorff distance between shapes, except that we compute the sum of closest distances, instead of their maxima. The summation operation makes the loss function differentiable w.r.t. the displaced points, and encourages the displaced point set to match the target tightly.   

In addition, we also compute a density loss. 
For each point $p$ in target point set $Y$, we define local density measures 
w.r.t. $Y$ and $\hat{Y}$, respectively, using two k-D vectors ($k = 8$ by default):
	\begin{eqnarray*}
		&[ \ d\big(p, N_1(Y, p) \big)\ \ \  d\big(p, N_2(Y, p) \big) \ \ \ ... \ \ \  d\big(p, N_k(Y, p) \big)  \  ] , \\
		&[ \ d\big(p, N_1(\hat{Y}, p) \big)\ \ \  d\big(p, N_2(\hat{Y}, p) \big) \ \ \ ... \ \ \  d\big(p, N_k(\hat{Y}, p) \big)  \  ].
		\label{eq:density-vector}
	\end{eqnarray*}
Here we denote $N_i(Y, p)$  as the  $i$-th closest point to $p$ from the same target point set $Y$, and $N_i( \hat{Y}, p)$  is the  $i$-th closest point to $p$ from the predicted point set $\hat{Y}$.

These two k-D vectors encode density measures of $Y$ and $\hat{Y}$ in small neighborhoods for each point $p\in Y$.
The density of the predicted point set $\hat{Y}$ resembles the density of target set $Y$, if and only if the density vectors of $\hat{Y}$ are similar to that of $Y$.
Therefore, the density loss is defined as the integration of distances between density vectors of $\hat{Y}$ and $Y$ over all points in $Y$:
\begin{eqnarray*}
	L_\text{density}(\hat{Y}, Y)  = \frac{1}{k} \sum\limits_{p\in{Y}}\sum\limits_{i=1}^k \big\vert{ d \big(p,N_i[Y,p] \big) - d\big(p, N_i[\hat{Y},p] \big) }\big\vert .
	\label{eq:density}
\end{eqnarray*}

With a setting of single X-to-Y transformation network, the geometric loss function is then as follows:
\begin{eqnarray*}
	L_{X \rightarrow Y}(\mathcal{D}) = \sum\limits_{\{X,Y\} \in \mathcal{D}} \Big( \ L_\text{shape}(\hat{Y}, Y) \ \  +  \ \  \lambda L_\text{density}(\hat{Y}, {Y}) \ \Big),
	\label{eq:single1}
\end{eqnarray*}
and similarily to the other direction:
\begin{eqnarray*}
	L_{Y \rightarrow X}(\mathcal{D}) = \sum\limits_{\{X,Y\} \in \mathcal{D}} \Big( \ L_\text{shape}(\hat{X}, X) \ \  +  \ \  \lambda L_\text{density}(\hat{X}, {X}) \ \Big),
	\label{eq:single2}
\end{eqnarray*}
where $\mathcal{D}$ denotes our training set, with a weight $\lambda = 1$ by default. 

\subsection{Cross Regularization}
\label{sec:regularization}

We couple the transformations $X$-to-$Y$ and $Y$-to-$X$  by a cross regularization over their displacement vectors $\mathcal{I}_X$ and $\mathcal{I}_Y$.
The key observation is that 
when $\mathcal{I}_X$ and $\mathcal{I}_Y$ are encouraged to be parallel to each other,  the two transformations can be mutually enhanced, with a more uniform distribution of the displacement mapping.

The regularization term maximizes the parallelism between $\mathcal{I}_X$ and $\mathcal{I}_Y$, without having paired displacements.
For each point $p \in X$, or  each point $q \in Y$, the displacements are associated with 6D vectors 
$[p , \ p+\mathcal{I}_X(p)]$ or  $[q+\mathcal{I}_Y(q), \ q ]$, respectively. The regularization works in 6D  in a similar manner as computing $L_\text{shape}$. That is: 
\begin{eqnarray*}
	L_\text{reg}(X, Y) &=& \sum\limits_{p \in X} \underset{q\in Y}{\min} \ d([p , \ p+\mathcal{I}_X(p)], \ \ [q+\mathcal{I}_Y(q), \ q ] )  \\
	&+& \sum\limits_{q \in Y} \underset{p\in X}{\min} \ d([p , \ p+\mathcal{I}_X(p)], \ \ [q+\mathcal{I}_Y(q), \ q ]). 
	\label{eq:reg}
\end{eqnarray*}

Minimizing $L_\text{reg}$ in 6D results in maximizing the parallelism between  3D displacement vectors with two opposite directions from bidirectional transformations.
See Fig.~\ref{fig:catdog}(c) that depicts the enhancement on a 2D toy example by adding cross regularization. More elaborated evaluation on the results is provided in Section~\ref{sec:results}. 

Given the displacement regularization that couples the transformations $X$-to-$Y$ and $Y$-to-$X$,  the network is trained with a loss function that sums three terms:
\begin{eqnarray}
	 L_{X \rightarrow Y}(\mathcal{D}) + L_{Y \rightarrow X  }(\mathcal{D}) + \mu \sum\limits_{\{X,Y\} \in \mathcal{D}} L_\text{reg}(X, Y),
	\label{eq:dual}
\end{eqnarray}
with a balancing parameter $\mu$ set as 0.1 by default.  

We minimize the loss~\eqref{eq:dual} with an Adam optimizer. The learning rate is set as 1e-3 and decays to 1e-4 at discrete intervals during training.

\begin{figure*}[th!]
	\centering
	\includegraphics[width=0.92\linewidth]{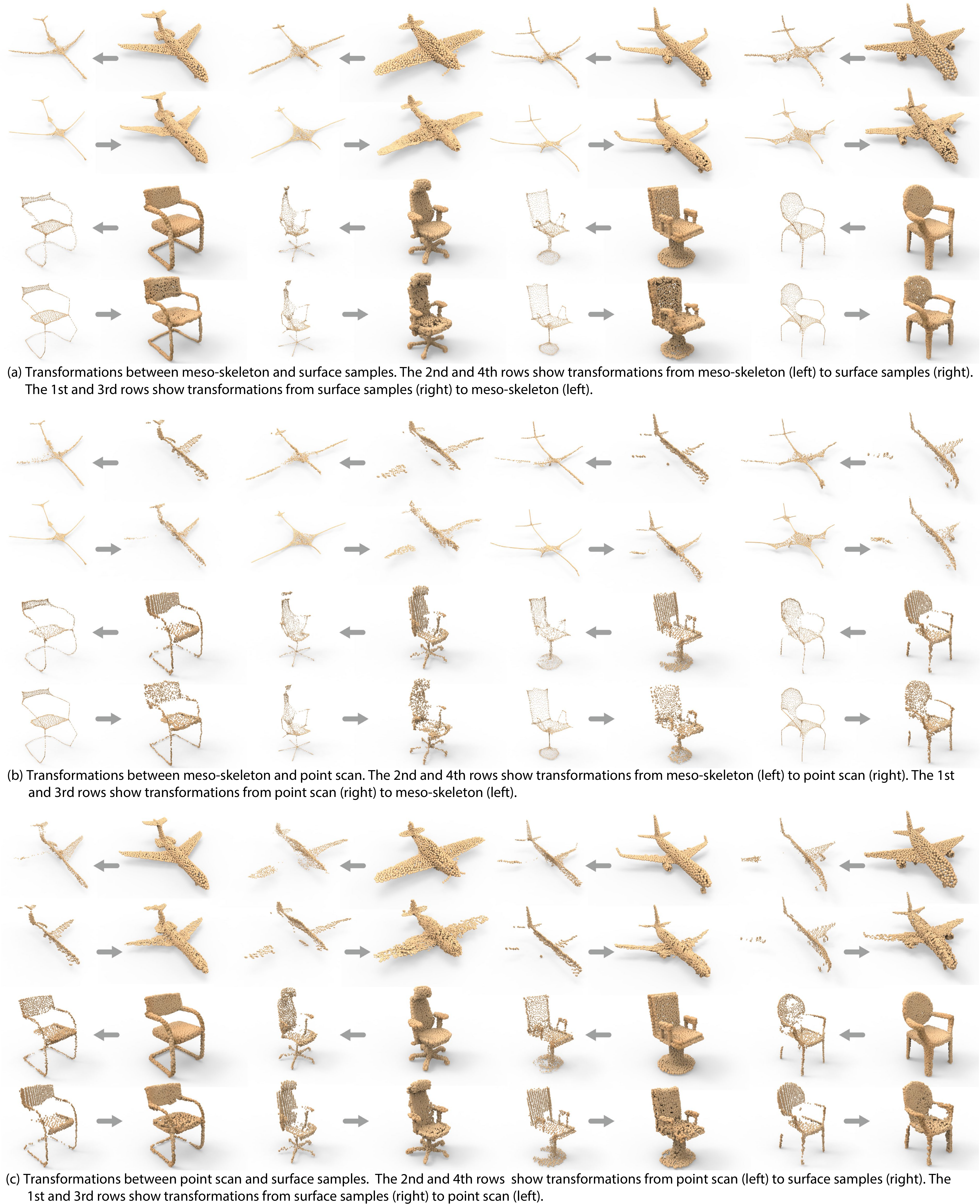}
	\caption{A gallery of point set transformations among meso-skeletons, shape surfaces, and single-view point scans via our network P2P-NET. Note that, to obtain the transformed surface point samples, we feed the same input eight times to the network and integrate the network outputs to produce a dense point set.}
	\label{fig:SSStransform}
\end{figure*}

\section{Experimental results and applications}
\label{sec:results}

We conduct experiments to demonstrate the capability of P2P-NET in learning geometric transforms between point sets in various domains.
Throughout the experiments, the network was trained using different datasets and for different domain pairs separately, but always with the same default 
network settings as described in Section~\ref{sec:method} and Appendix~\ref{sec:append_detail}. There is no hyperparameter or architecture tuning for any specific 
domains or any specific datasets. All the results are presented without any post-processing.

\subsection{Meso-skeleton, surface, and single-view point scan}
\label{sec:skeletonshape}

In many cases, a mapping from one domain to another is easy, but the inverse is a lot more difficult.
For example, synthesizing point scans from 3D shapes is easy, but surface completion is hard.
Skeleton extraction from 3D shapes may have been considered as a solved problem~\cite{Tag16}, but synthesizing shape surfaces 
from skeletons is an unresolved challenge. 
Our network is able to learn to solve ill-posed inverse mapping problems (e.g., skeleton-to-surface) by using training data synthesized 
by an algorithm designed for the easier transform (e.g., surface-to-skeleton). In this section, we demonstrate transformations
among meso-skeletons, surface samples, and single-view point scans using P2P-NET.

Given a set of 3D shapes, we convert them to surface samples with Poisson disk sampling~\cite{corsini2012efficient}. 
By taking the surface samples as input, meso-skeletons of the shapes are obtained using a contraction-based approach~\cite{cao2010point}.
To show the robustness of our network to shape occlusion, we also synthesize single view point scans with a Kinect simulator~\cite{bohg2014robot,gschwandtner2011blensor} applied to 3D shapes. 
We use the chair and airplane datasets of ModelNet40~\cite{wu20153d} as original 3D shapes, and sample each point set to the size of 2,048. The chair dataset contains 889 training and 100 test examples, while the airplane dataset contains 626 training and 100 test examples. 

With the synthesized surface samples, meso-skeletons, and single-view point scans, we tested our method on three pairs of transformations among the three different types of point sets, i.e., meso-skeleton vs.~surface, meso-skeleton vs.~point scan, and point scan vs.~surface.
In Fig.~\ref{fig:SSStransform}, the visual results of the three pairs of transformations are provided with eight distinctive examples chosen from the test set. 
Note that, in order to obtain transformed surface point samples, we feed the same input in eight passes to P2P-NET and integrate the network outputs to produce a dense final result.
To obtain point scans or meso-skeletons, we only feed the input once to the network. 

To convey that our network is able to learn a shape transform, 
we show the closest training examples retrieved for the inputs over the eight test examples in Appendix~\ref{sec:append_closetrain}.  We also provide quantitative evaluations for the three pairs of transformations in Section~\ref{sec:evaluation}.

\begin{figure}[!t]
	\centering
	\includegraphics[width=0.99\linewidth]{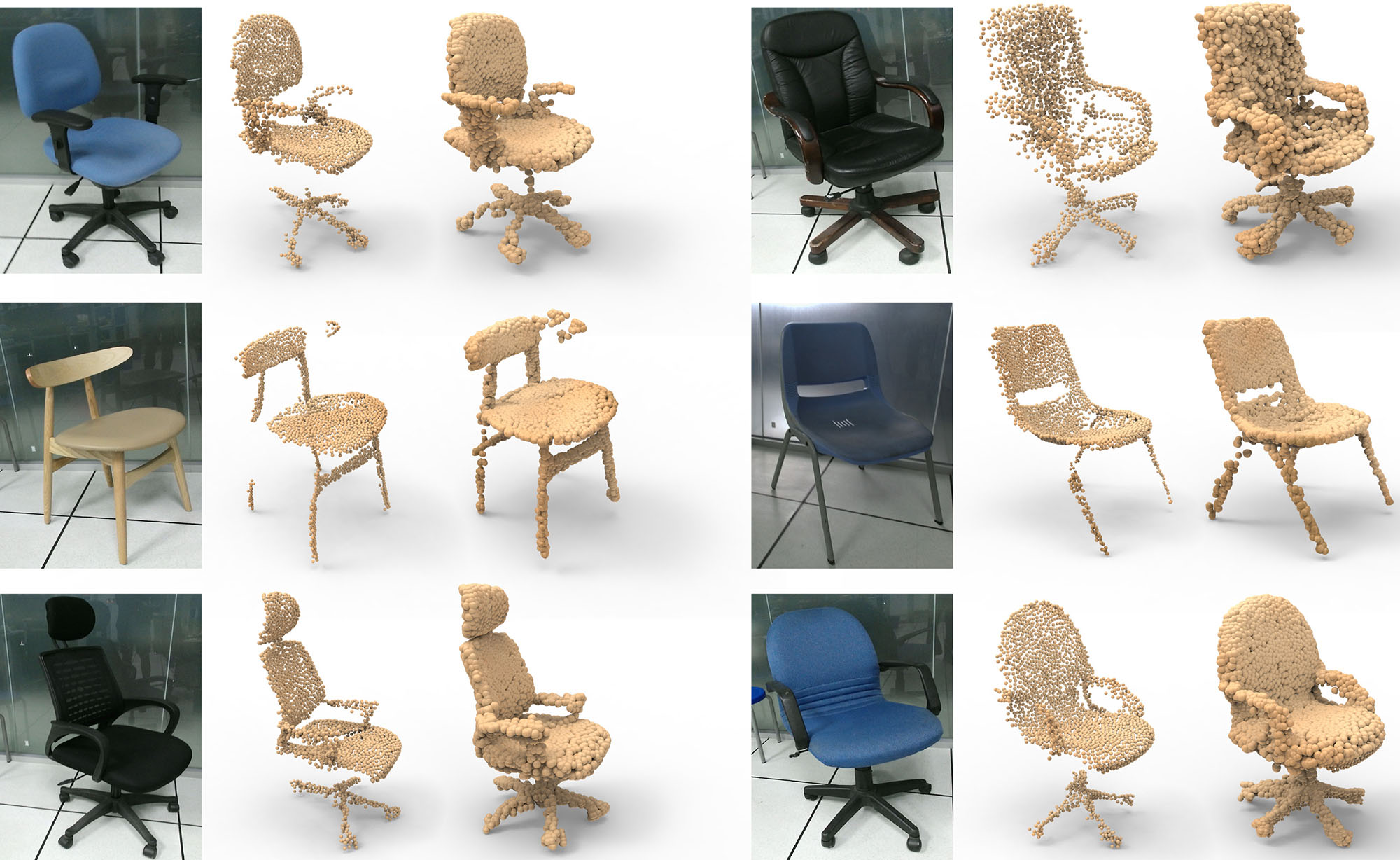}
	\caption{\rz{Testing P2P-NET on real chair scans (middle) captured by Kinect v2. The completed point cloud is shown on the right.}} 
	\label{fig:realScan}
\end{figure}

Fig.~\ref{fig:displace_visul} visualizes point-wise displacements produced by P2P-NET to offer a glimpse of
what the network learned. We re-emphasize that the network was not trained on any point-wise mapping between paired shapes nor with any 
displacement vectors. Yet, it appears that the learned displacements are well-localized and reflect what a properly devised transformation
algorithm would produce.

\rz{In Fig.~\ref{fig:realScan}, we show that the trained P2P-NET is capable of converting real point scans of chairs
captured by a Kinect v2 to complete shapes. Note that during the capture, the Kinect sensor was placed to roughly 
align with the camera view used in data synthesis.}

\begin{figure}[!t]
	\centering
	\includegraphics[width=0.95\linewidth]{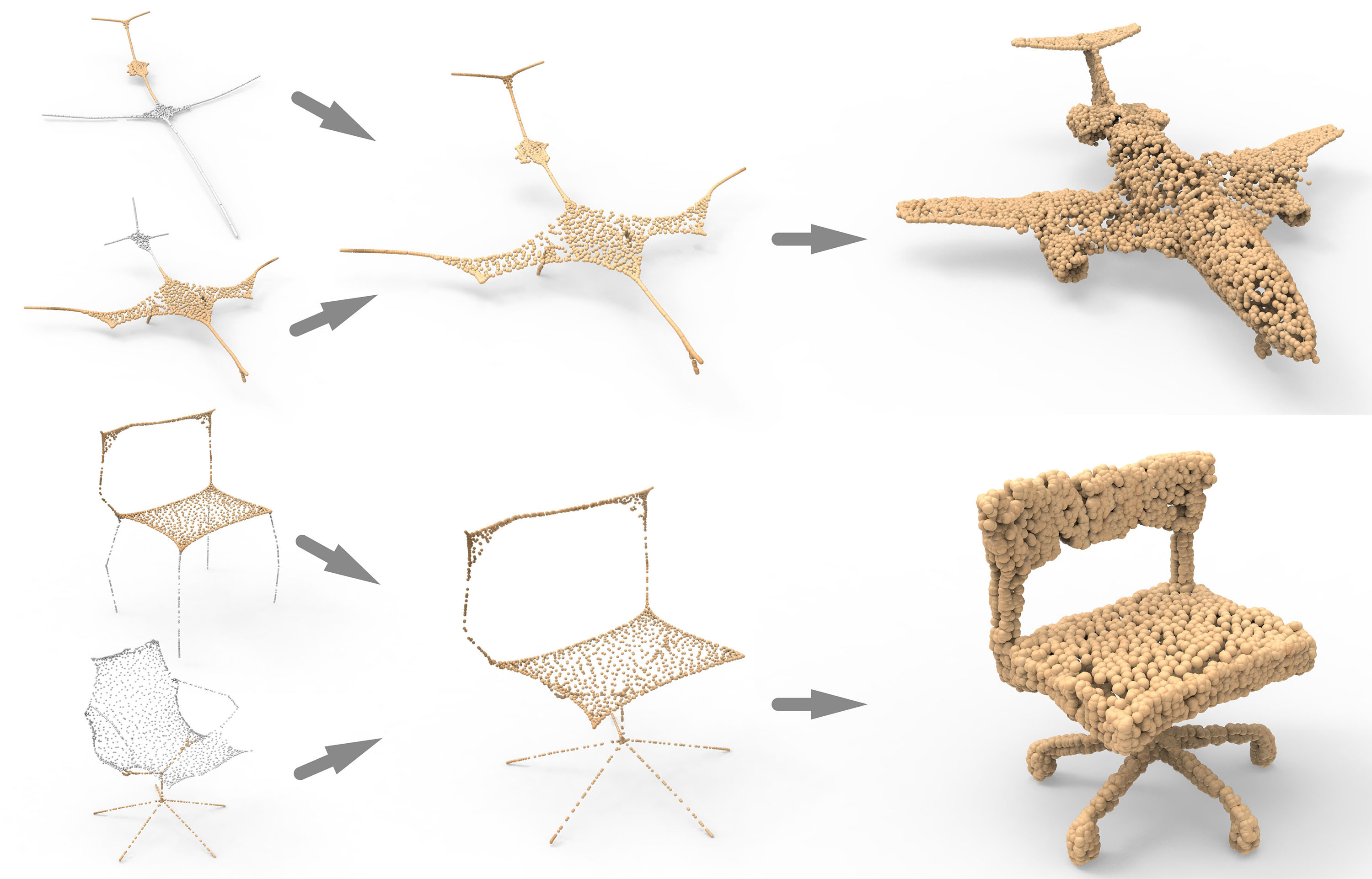}
	\caption{After editing and combining the point sets of meso-skeletons, our network is able to generate new shapes (right) from the new meso-skeletons. }
	\label{fig:shapeedit}
\end{figure}

The results shown in Fig.~\ref{fig:SSStransform} demonstrate the potential of P2P-NET for possible applications.
In Fig.~\ref{fig:shapeedit}, we show such an example for shape editing and synthesis. After combining the 
meso-skeletons of different shapes into a new meso-skeleton, our network can convert the synthesized meso-skeleton 
into a new point-set shape. Moreover, the result of transforming point scans to surface samples offers the promise of
applying P2P-NET for scan completion. To extend P2P-NET to a full-fledged scan completion network, one would 
require a multi-view assembly of the network or adding view prediction and rotation layers, which are out of the 
scope of this paper. 

\begin{figure*}[t!]
	\centering
	\includegraphics[width=0.9\linewidth]{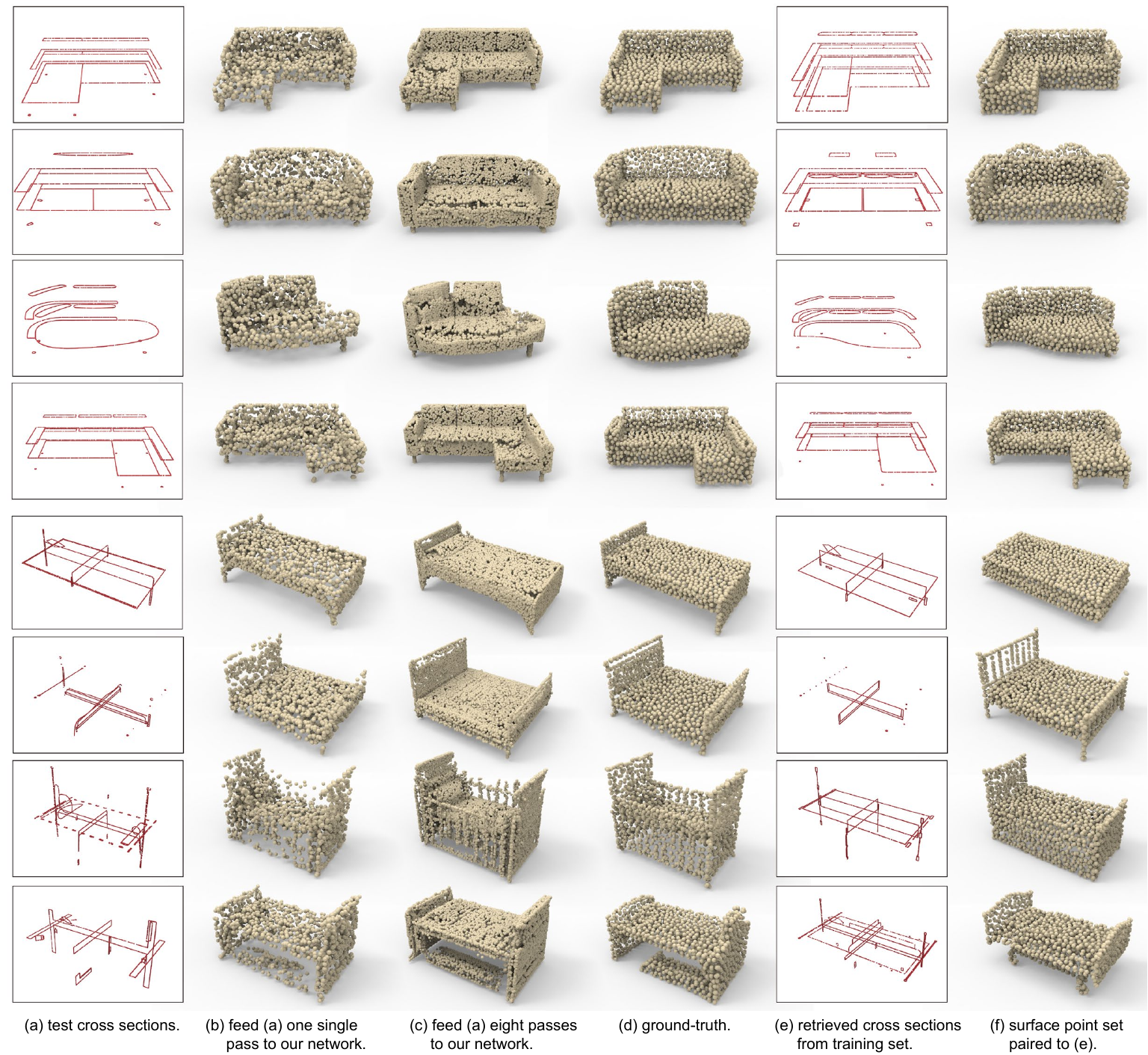}
	\caption{Transformations from 2D cross-sectional profiles (a) to 3D object surfaces (b) and (c). In addition to ground-truths (d), we also provide the closest 2D cross-sectional profiles (e) retrieved from the training set, and their corresponding surface point sets (f).}
	\label{fig:crosssection}
\end{figure*}

\subsection{2D cross-sectional profiles and 3D shapes}
\label{sec:crossection}

\begin{figure*}[t!]
	\centering
	\includegraphics[width=0.85\linewidth]{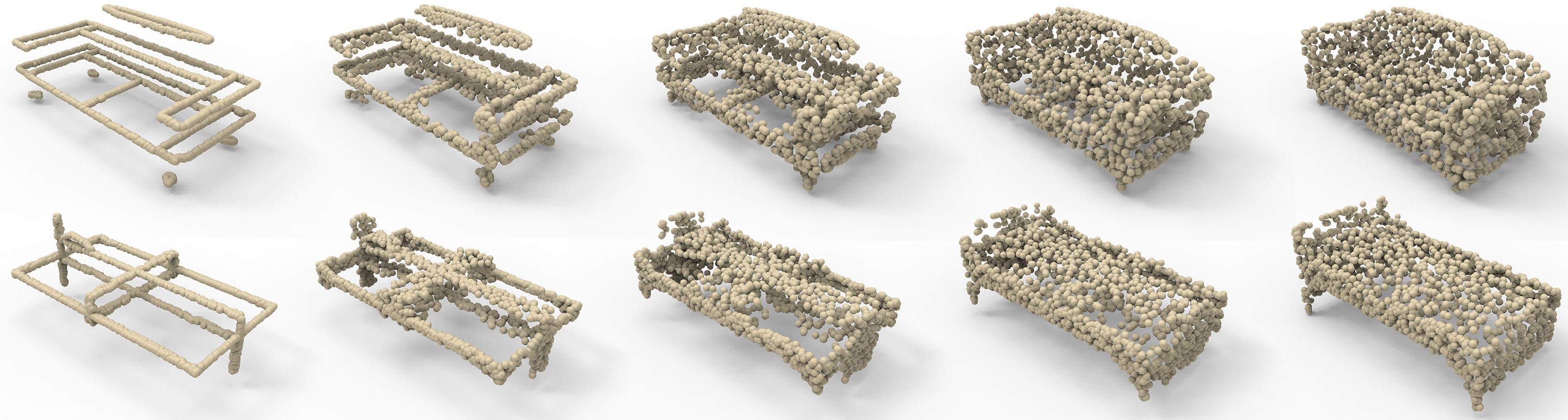}
	\caption{Visualization of point displacements learned by P2P-NET, which transform cross-sectional profiles into surface samples.  
We scaled the displacements, from left to right, by factors of 0.05, 0.25, 0.5, 0.75, 1.0, respectively, to obtain a morphing sequence.}
	\label{fig:bedsofa_morph}
\end{figure*}

Planar cross-sectional profiles are widely used in computer-aided design and geometric modeling.  
Transforming from 2D cross sections to 3D shapes is an interesting test for our neural network.
For this experiment, we use the sofa and bed datasets of ModelNet40. 
The sofa dataset contains 680 training and 100 test examples. 
Similarly, in the bed dataset, there are 515 training and 100 test examples.  
We cut each sofa with four parallel planes to obtain four parallel cross sections, and cut each bed with 
three orthogonal planes to obtain three orthogonal cross sections. We sample each set of cross sections uniformly to acquire a point set consisting of 2,048 points. 
Each point set of cross sections is then paired with the point set of mesh samples of the same sofa or bed object. 

We visualize the results obtained on eight typical test examples of the sofa and bed in Fig.~\ref{fig:crosssection}. 
We can observe that the transformation results obtained with feeding the network a single input pass exhibit non-uniformity and missing regions.
However, after feeding the input over eight passes and integrating the network outputs, the resulting dense point sets are complete and smooth overall. 
This demonstrates the stability of the transformation prediction by P2P-NET.

It is also interesting to observe that, in some cases, the dense outputs produced by multiple passes of P2P-NET can better convey
shape details than the ground truth data, e.g., see column (c) in Fig.~\ref{fig:crosssection} for the long pillow in the
fifth row and the slats in the crib in the second to last row, in contrast to their counterparts in column (d). One reason
is that the ground truth is only at $1/8$ of the resolution, compared to the dense results. On the other hand, the level
of surface details produced by the network are not copied from the training set, since all training data are at a low
resolution of 2,048 points which do not well reflect the surface details. The produced details should be attributed to the
point transforms learned by P2P-NET.


To further demonstrate that the network has learned a proper transform, we retrieved the closest training cross sections with the test cross sections as query inputs,
and show the retrieved cross sections and their paired sampled meshes in Fig.~\ref{fig:crosssection}. The retrieval was carried out using the distance measure:
\begin{eqnarray*}
	D_\text{retrieve}(P,Q) = \sum\limits_{p \in P} \underset{q\in Q}{\min}  \left\lVert p-q \right\rVert   + \sum\limits_{q \in Q} \underset{p\in P}{\min} \left\lVert p-q \right\rVert,
	\label{eq:retrieve}
\end{eqnarray*}
where $P$ and $Q$ are two point sets of cross sections.

We can observe that the retrieved cross sections are generally not close to the queries. This is clearly evident in the last three
rows of Fig.~\ref{fig:crosssection}. Admittedly, it is far from trivial to come up with an accurate 
similarity distance measure for cross-sectional profiles. To confirm that the retrieved results are reasonable, we have manually
examined all the training examples in the dataset and found no other cross sections to be visually closer to those shown
in the figure.


In Fig.~\ref{fig:bedsofa_morph}, we visualize the displacement vectors learned for the current transform. Since mappings from
cross-sectional profiles to 3D shapes are much less predictable and coherent, instead of showing the displacement vectors
explicitly like in Fig.~\ref{fig:displace_visul}, we show a morphing sequence following the displacements.

\rz{What is common between skeleton-to-shape and profile-to-shape transforms is that one domain has an {\em easy-to-edit\/} 
shape abstraction. It is quite common to perform user edits on skeletons and curve profiles. After that, it would be quite desirable to 
be able to directly convert the edited shape abstractions to whole shapes.\/}
Like the example shown in Fig.~\ref{fig:shapeedit}, we also experimented with editing 2D cross-sections
and then transforming the edits to 3D point-set shapes using P2P-NET. A visual result showing an interpolating sequence is provided in
Fig.~\ref{fig:teaser}. This further demonstrates the potential of our network in shape synthesis applications.


\subsection{Quantitative evaluation}
\label{sec:evaluation}

\begin{table*}[t]
	\centering
	\resizebox{0.85\linewidth}{!}{
		\begin{tabular}{|c|l|l|l|r|r|r|c|r|r|r|l|r|r|r|}
\hline
\multirow{2}{*}{Dataset}  & \multicolumn{1}{c|}{\multirow{2}{*}{Source}} & \multicolumn{1}{c|}{\multirow{2}{*}{Target}} &  & \multicolumn{3}{c|}{mean of separation rate}                                                    &  & \multicolumn{3}{c|}{mean of curvature diff.}                                                    &                       & \multicolumn{3}{c|}{mean of normal diff.}                                                       \\ \cline{4-15} 
                          & \multicolumn{1}{c|}{}                        & \multicolumn{1}{c|}{}                        &  & \multicolumn{1}{c|}{ns-rg-} & \multicolumn{1}{c|}{ns+rg-} & \multicolumn{1}{c|}{ns+rg+} &  & \multicolumn{1}{c|}{ns-rg-} & \multicolumn{1}{c|}{ns+rg-} & \multicolumn{1}{c|}{ns+rg+} & \multicolumn{1}{c|}{} & \multicolumn{1}{c|}{ns-rg-} & \multicolumn{1}{c|}{ns+rg-} & \multicolumn{1}{c|}{ns+rg+} \\ \hline
\multirow{6}{*}{airplane} & skeleton                                     & surface                                      &  & 1.4\%                       & \textbf{0.6\%}              & \textbf{0.6\%}              &  & 0.084                       & 0.063                       & \textbf{0.062}              &                       & 0.575                       & 0.390                       & \textbf{0.389}              \\ \cline{2-15} 
                          & surface                                      & skeleton                                     &  & \textbf{0.3\%}              & 0.4\%                       & \textbf{0.3\%}              &  & 0.076                       & 0.076                       & \textbf{0.075}              &                       & \multicolumn{1}{c|}{-}      & \multicolumn{1}{c|}{-}      & \multicolumn{1}{c|}{-}      \\ \cline{2-15} 
                          & scan                                         & skeleton                                     &  & 2.1\%                       & \textbf{2.0\%}              & \textbf{2.0\%}              &  & 0.068                       & 0.068                       & \textbf{0.066}              &                       & \multicolumn{1}{c|}{-}      & \multicolumn{1}{c|}{-}      & \multicolumn{1}{c|}{-}      \\ \cline{2-15} 
                          & skeleton                                     & scan                                         &  & 2.9\%                       & \textbf{2.3\%}              & 2.4\%                       &  & \textbf{0.051}              & 0.052                       & 0.052                       &                       & 0.677                       & 0.617                       & \textbf{0.601}              \\ \cline{2-15} 
                          & scan                                         & surface                                      &  & 1.6\%                       & \textbf{1.2\%}              & 1.3\%                       &  & 0.074                       & 0.063                       & \textbf{0.061}              &                       & 0.495                       & 0.444                       & \textbf{0.418}              \\ \cline{2-15} 
                          & surface                                      & scan                                         &  & 1.3\%                       & \textbf{1.1\%}              & 1.3\%                       &  & \textbf{0.056}              & 0.057                       & 0.057                       &                       & 0.669                       & 0.670                      & \textbf{0.667}              \\ \hline
\multirow{6}{*}{chair}    & skeleton                                     & surface                                      &  & 12.0\%                      & \textbf{7.5\%}              & \textbf{7.5\%}              &  & 0.096                       & 0.082                       & \textbf{0.080}              &                       & 0.686                       & 0.620                       & \textbf{0.617}              \\ \cline{2-15} 
                          & surface                                      & skeleton                                     &  & 5.3\%                       & 5.3\%                       & 5.3\%                       &  & 0.061                       & 0.061                       & \textbf{0.060}              &                       & \multicolumn{1}{c|}{-}      & \multicolumn{1}{c|}{-}      & \multicolumn{1}{c|}{-}      \\ \cline{2-15} 
                          & scan                                         & skeleton                                     &  & 15.5\%                      & 15.8\%                      & \textbf{15.4\%}             &  & 0.053                       & 0.053                       & \textbf{0.051}              &                       & \multicolumn{1}{c|}{-}      & \multicolumn{1}{c|}{-}      & \multicolumn{1}{c|}{-}      \\ \cline{2-15} 
                          & skeleton                                     & scan                                         &  & 18.8\%                      & 17.6\%                      & \textbf{17.5\%}             &  & \textbf{0.052}              & 0.054                       & \textbf{0.052}              &                       & 0.590                       & 0.584                       & \textbf{0.562}              \\ \cline{2-15} 
                          & scan                                         & surface                                      &  & 10.9\%                      & \textbf{6.6\%}              & 6.7\%                       &  & 0.092                       & \textbf{0.083}              & 0.084                       &                       & 0.613                       & 0.557                       & \textbf{0.553}              \\ \cline{2-15} 
                          & surface                                      & scan                                         &  & 5.2\%                       & 5.2\%                       & \textbf{4.9\%}              &  & 0.057                       & 0.057                       & \textbf{0.056}              &                       & 0.553                       & 0.556                       & \textbf{0.552}              \\ \hline
\multirow{2}{*}{sofa}     & cross sec.                                   & surface                                      &  & 22.0\%                      & \textbf{9.8\%}              & 10.8\%                      &  & 0.084                       & 0.066                       & \textbf{0.065}              &                       & 0.626                       & \textbf{0.457}              & 0.458                       \\ \cline{2-15} 
                          & surface                                      & cross sec.                                   &  & 9.6\%                       & 9.2\%                       & \textbf{8.9\%}              &  & \textbf{0.059}              & 0.060                       & \textbf{0.059}              &                       & \multicolumn{1}{c|}{-}      & \multicolumn{1}{c|}{-}      & \multicolumn{1}{c|}{-}      \\ \hline
\multirow{2}{*}{bed}      & cross sec.                                   & surface                                      &  & 14.5\%                      & 3.0\%                       & \textbf{2.9\%}              &  & 0.084                       & \textbf{0.056}              & \textbf{0.056}              &                       & 0.544                       & \textbf{0.380}              & \textbf{0.380}              \\ \cline{2-15} 
                          & surface                                      & cross sec.                                   &  & \textbf{11.0\%}             & 11.8\%                      & 11.5\%                      &  & 0.058                       & 0.058                       & 0.058                       &                       & \multicolumn{1}{c|}{-}      & \multicolumn{1}{c|}{-}      & \multicolumn{1}{c|}{-}      \\ \hline
\end{tabular}
	}
	\caption{Quantitative evaluation of our network with different settings and error metrics. In the head of the table, `ns' stands for noise augmentation, `rg' stands for cross regularization, and `+/-' indicates enable/disable. }
	\label{table:eva}
\end{table*}

We evaluated the performance of our P2P-NET quantitatively on the four datasets used in Sections~\ref{sec:skeletonshape} and~\ref{sec:crossection}.
For the purpose of measuring the errors, the original shapes and their point sets were normalized, so that the diagonal lengths of their bounding boxes are equal to 1. It should be remembered that all the training and test point sets are sampled to the size of 2,048.  The performances were measured with three error metrics, as detailed below.

\vspace{-5pt}

\paragraph{\bf{Point separation rate.}}
Given a predicted point set and the ground-truth point set,  each point searches for its closest point from the opposite set. If the distance from a point $p$ to its closest point $q$ in the opposite set is greater than 0.02, we consider the point $p$ as a separated point.  We call the percentage of separated points among all points in the two sets, the separation rate. We compute a separation rate for every test example and report the mean in Table~\ref{table:eva}.

\vspace{-5pt}

\paragraph{\bf{Curvature difference.}}
We estimate a curvature indicator for each point $p$ as $\lambda_0/(\lambda_0+ \lambda_1+ \lambda_2)$, 
where $\lambda_0 \leq \lambda_1 \leq \lambda_2$ are the eigenvalues of a $3\times 3$ covariance matrix of a local point patch around point $p$.
The size of the local patch is 0.3\% of the size of whole point set.
For each point $p$, we compute the absolute difference of curvature indicator with its closest point $q$ in the opposite point set.  
Finally, we compute the mean difference of all points.

\vspace{-5pt}

\paragraph{\bf{Normal difference.}}
Similar to the curvature indicator, we estimate a PCA normal for each point, and measure the radian of the angle between the normal of a point $p$ and the normal of its closest point $q$ in the opposite set. We compute the mean radian of all points.

For each pair of transformations on each dataset, we trained the networks for 200 epoches on a Nvidia Titan Xp GPU that takes approximately 5$\sim$8 hours to finish the process. During the testing phase, when the target point set is not from surface samples, we feed the source point set in one single pass to the network and obtain output point sets of size 2,048. When the target point set is from surface samples, we feed the source point set in eight passes to the network to obtain an integrated dense output of size 16,384.  Four pairs of transformations and three different settings of the network were tested and reported in Table~\ref{table:eva}.  
The three different network settings are: no noise augmentation and no cross regularization (ns-rg-),  no noise augmentation but with cross regularization (ns-rg+), and the setting
with both options on (ns+rg+).  

Note that the separation rate measures the tightness of the match between the predicted point set and the ground-truth point set.
The smaller the value is, the tighter the match. The results in Table~\ref{table:eva} show that noise augmentation helps to reduce the separation rate, 
i.e., to make the predicted points to match more tightly to the ground-truth point sets.  
Adding the cross-regularization does not further reduce the separation rate. 
However, as shown in Table~\ref{table:eva}, adding the cross-regularization term does achieve lower error rates overall, in terms of curvature difference and normal difference. Since these two measures reflect how well local geometric properties of the point set are preserved, the quantitative results demonstrate that cross-regularization is effective in enhancing the local geometric properties of the predicted point set.  

\section{Discussion, limitation, and future work}
\label{sec:future}
  
By design, P2P-NET is a general-purpose point-to-point displacement network, in that no parts of the network are 
tailor-made to specific transformation tasks. Moreover, we do not alter the network architecture, 
when dealing with different pairs of transformation domains.
The network is trained to map point sets from one domain to another, where the point sets can be in 2D or 3D spaces. 
As we demonstrated, the mapping can also lift 2D profiles to 3D shapes; see Fig.~\ref{fig:crosssection}.
Since the mapping is applied in a feature space, the learned transform is agnostic to the dimensionality of the point sets. 
Interestingly, the point displacements, which are one-to-one, are learned without training data on point-wise 
mapping or displacement vectors. 
All we provide are pairs of point-set shapes, which may even have different cardinalities. 


\begin{figure}[t!]
	\centering
	\includegraphics[width=\linewidth]{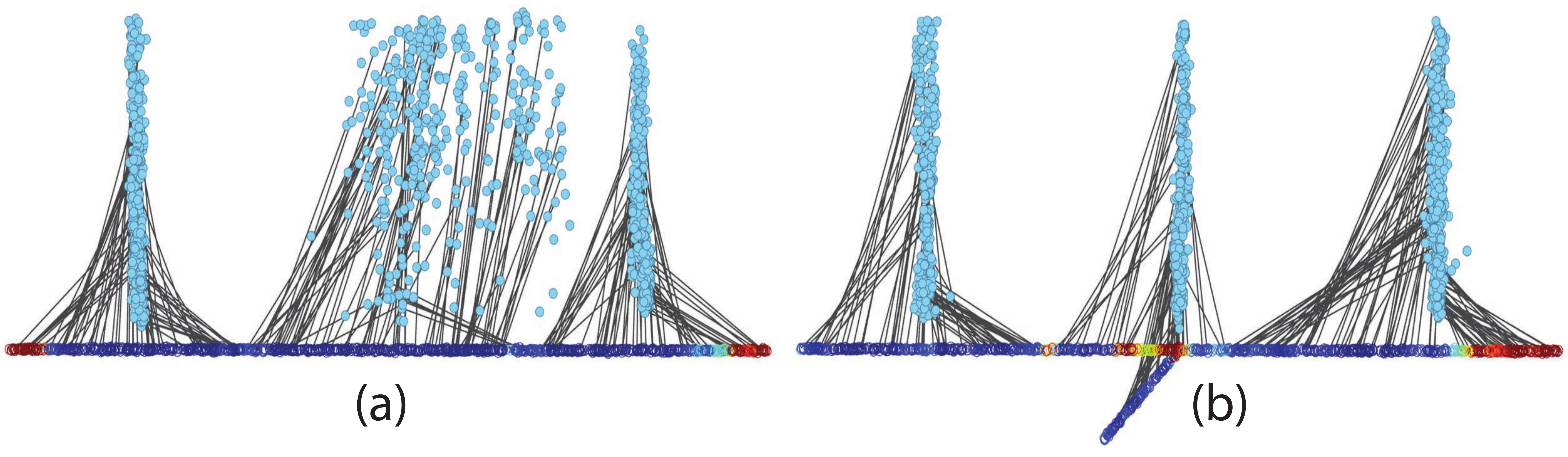}
	\caption{\rz{P2P-NET cannot be properly trained to map a long horizontal line to three vertical bars, since points near the mid-section possess similar point features. Our method fails to associate points with similar features with dissimilar displacements to clearly form the middle bar (a). However, a small added protrusion (b) can serve to disambiguate these point features, leading to different displacements to produce the middle bar. Colorings of the points reflect their features, after a 1D embedding using PCA.}}
	\label{fig:featureless}
\end{figure}

P2P-NET is bidirectional, which may be reminiscent of networks trained under cycle consistency~\cite{zhu2017unpaired,yi2017dualgan}. 
However, there is no cyclic consistency in P2P-NET; the bidirectionality is used to form a cross-regularization 
which exploits the two mappings to enhance the mapping distribution. 
\rz{There is an intriguing, and seemingly ``dual'', relation between P2P-NET and CycleGAN~\cite{zhu2017unpaired}. P2P-NET 
is trained on paired shapes, while CycleGAN learns from unpaired images. But in CycleGAN, there is pixel-to-pixel correspondence 
between the pair of training images; P2P-NET does not require point-to-point correspondence between the training point-sets.
CycleGAN is trained to learn how to transform pixel values, {\em in place\/}, while Point-NET is trained to displace points from 
one shape representation to another.}

\begin{figure}[t!]
	\centering
	\includegraphics[width=0.95\linewidth]{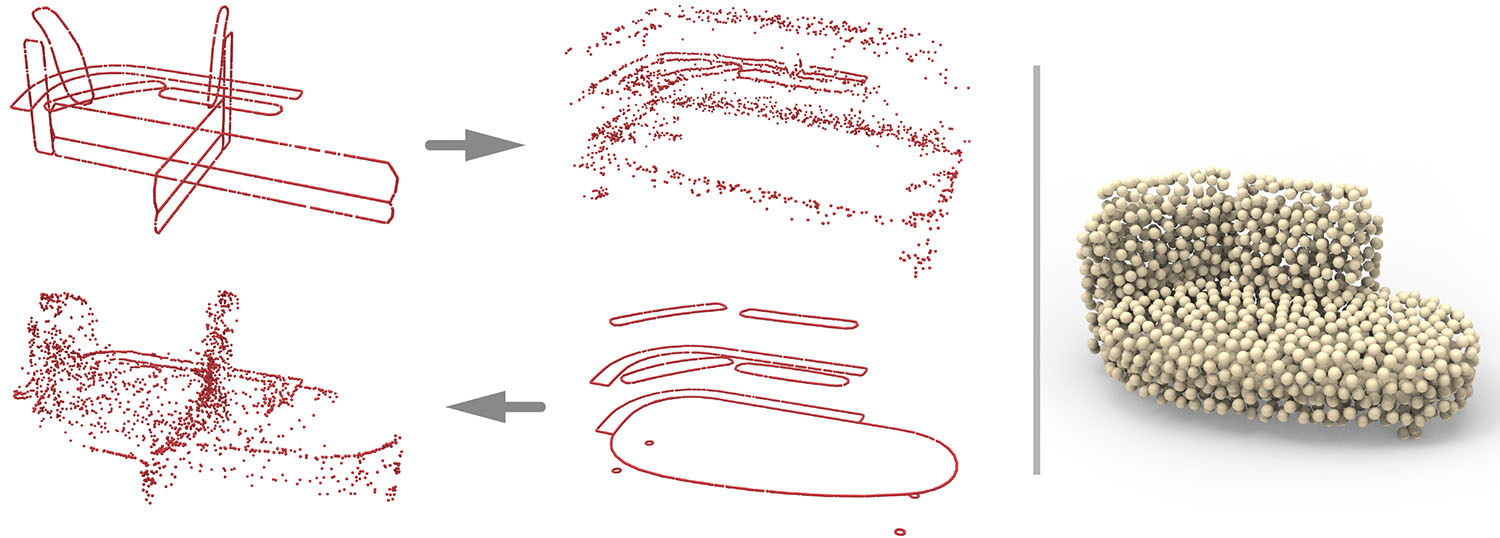}
	\caption{\rz{P2P-NET cannot be trained to produce highly structured point sets cleanly, e.g., to transform between orthogonal profiles and parallel profiles for the same 3D shape (sofa on the right).}}
	\label{fig:profiletrans}
\end{figure}

\rz{
We reiterate that our work is only a first attempt at designing a general-purpose shape transformation 
network. By no means should one expect P2P-NET to work effectively for all transformation tasks. The
network is inherently limited by its current architecture, training loss, and optimization scheme for the
network parameters. In what follows, we provide a non-exhaustive list of such limitations to explore the
behavior and limit of our method.


\begin{figure}[t!]
	\centering
	\includegraphics[width=0.85\linewidth]{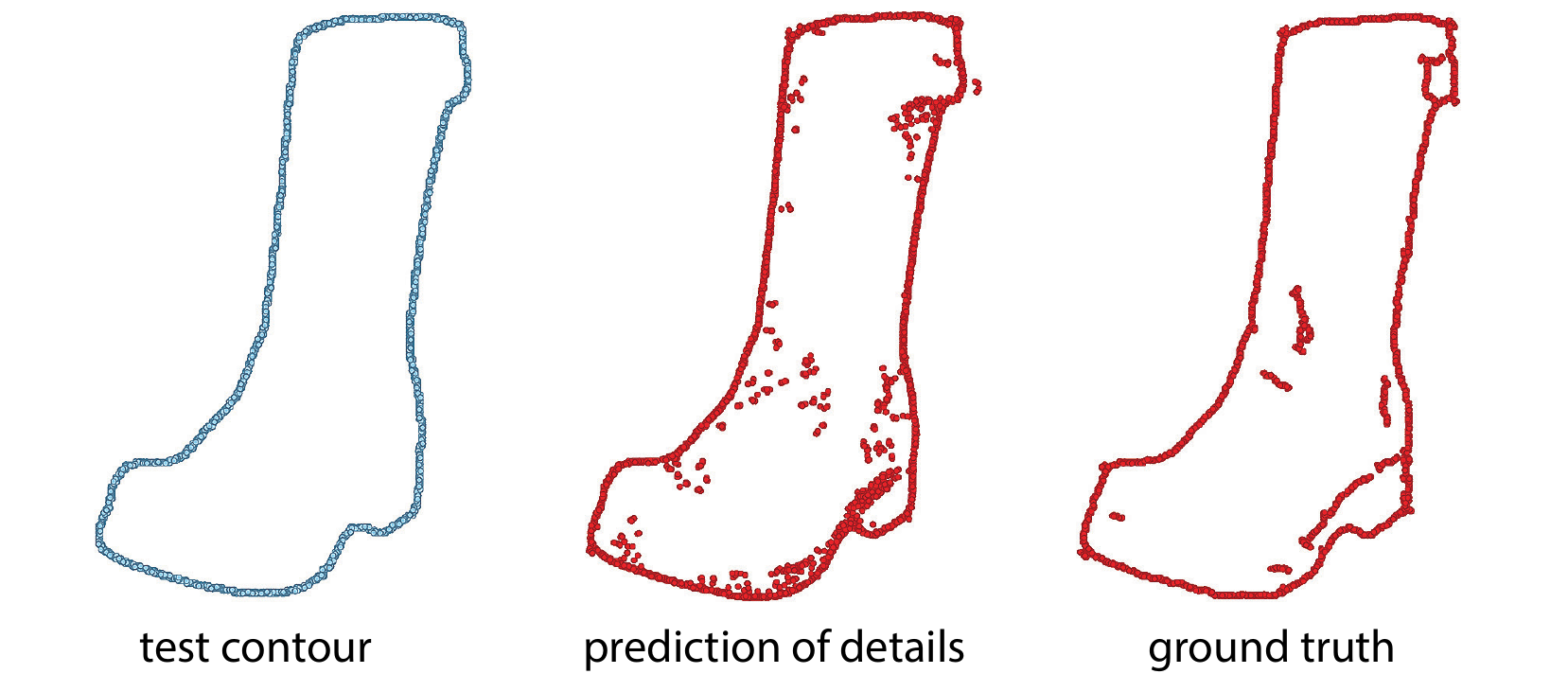}
	\caption{P2P-NET cannot be properly trained for non-deterministic \rz{point set transforms}, e.g., adding details to a shape contour.}
	\label{fig:boot}
\end{figure}

\vspace{-5pt}

\paragraph{\bf{Ambiguous feature-to-displacement mapping.}}
In the absence of point-to-point correspondences between the training source and target shapes, P2P-NET
must learn point set transforms implicitly. Architecturally, P2P-NET first turns the input points into PointNET++
features. It then learns to map these point features to displacement vectors to minimize the training loss. As a 
result, P2P-NET should be trained with examples, where the implicit relation between point features and 
displacements is {\em unambiguous\/}. The network should not be expected to learn to associate
points possessing similar features with {\em different\/} displacements.
In reality, however, ambiguous feature-to-displacement mappings may be unavoidable for many transformation 
tasks. They may be characteristic of an entire class of transformations or occur only for some shape 
pairs or only over a portion of the shapes. Any such case may potentially lead to failure cases by P2P-NET.

To provide a simple illustration, consider a 2D example of learning to map points
along a long horizontal line to three vertical bars, as shown in Fig.~\ref{fig:featureless}. We trained P2P-NET
using more than 1,000 examples of source and target pairs in random orientations and scales. However,
regardless of how many training examples we employed, the network still cannot map the mid-section
of the line to the middle bar, since points near the mid-section all possess similar features. As shown in (a),
P2P-NET could only learn to associate these points with similar displacements. To verify that the crux of
the problem is the ambiguity, we added a small protrusion under the source line, so that points near the
mid-section can be better distinguished by PointNET++ features. As can be seen in (b),  P2P-NET now
does a much better job of learning the proper transform.

%

\vspace{-5pt}

\paragraph{\bf{Ambiguous point transforms.}}
Some point set transforms may exhibit shape-level ambiguities, as shown in Fig.~\ref{fig:boot}. In this task, we 
learn displacements from a shape silhouette to its interior details. The training set contains more
than 1,000 examples of adding different details (via edge maps) to different boot shapes. To minimize the training loss, 
P2P-NET is only able to learn to displace to an {\em average\/} of the target points, leading to an erroneous outcome. 

%

\begin{figure}[!t]
	\centering
	\includegraphics[width=0.95\linewidth]{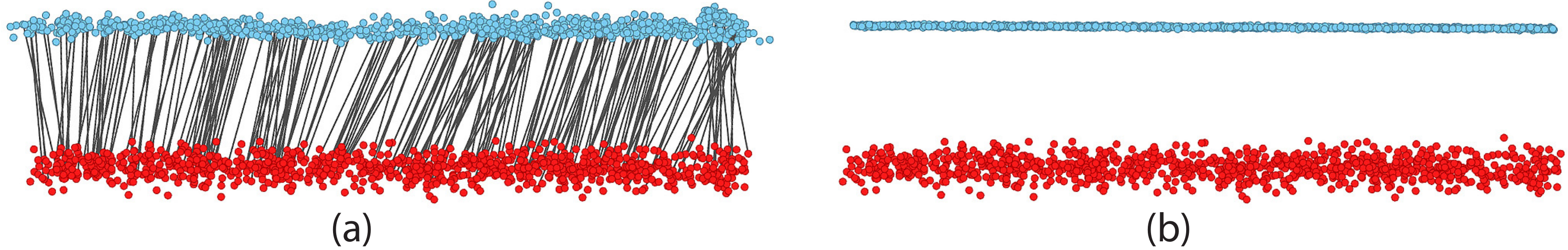}
	\includegraphics[width=0.95\linewidth]{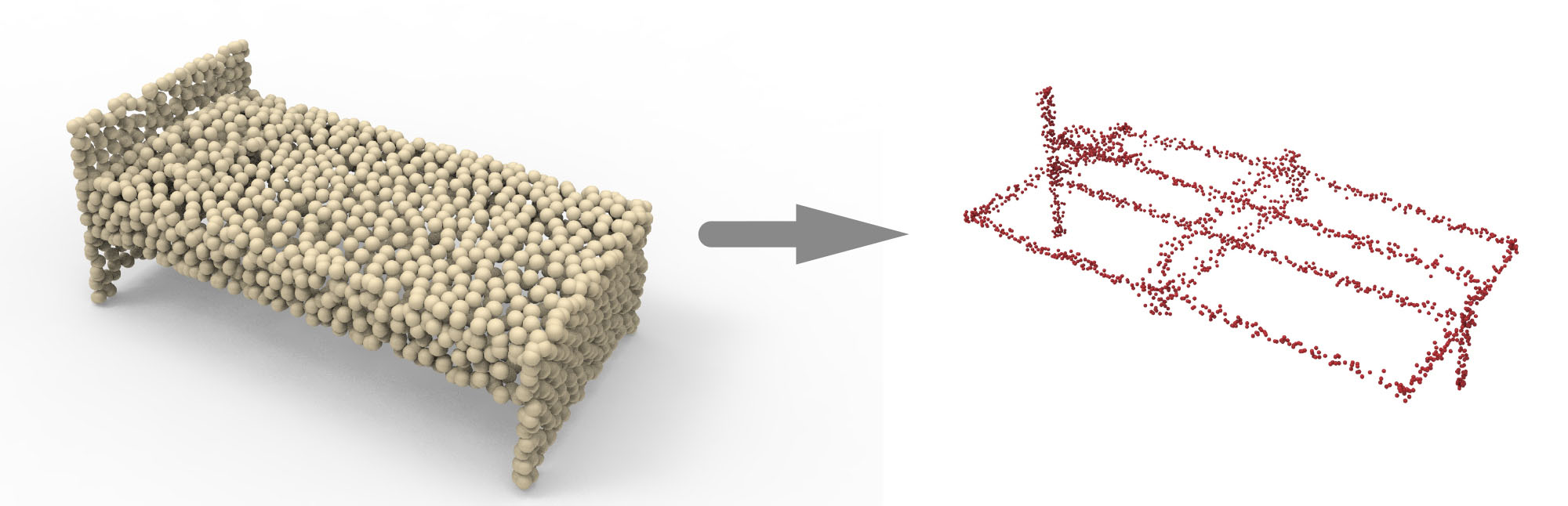}
	\caption{\rz{P2P-NET cannot be trained to accurately and cleanly predict thin structures. Top: from a noisy input (red), as shown in (a), in contrast to the ground truth (b).
			Bottom: from lower-resolution point set inputs.}}
	\label{fig:noisycross}
\end{figure}


%


\vspace{-5pt}

\paragraph{\bf{Decorrelation of displacements.}} 
The training loss adopted by P2P-NET is predominantly a point-to-shape distance measure. It does not account for intrinsic
properties of the input shape. This immediately implies that P2P-NET is generally unable to learn such properties, so as to
preserve them in the output point clouds. The point displacement vectors predicted by P2P-NET are not correlated or controlled by
the shape properties, since the network predicts a displacement vector for each point independently.
%
Fig.~\ref{fig:noisycross} shows that P2P-NET is unlikely to reproduce thin lines, when all the training data contain clean,
thin line structures.
Fig.~\ref{fig:profiletrans} shows that P2P-NET cannot be well trained to produce highly structured point sets, where the transform is
between orthogonal profiles and parallel profiles of the same shapes.
By the same token, P2P-NET is not part-aware, i.e., it is unlikely to preserve part structures of the input shapes. For example, it
cannot transform clean rooms into messy rooms, by displacing or adding point-set objects. Currently, only uncorrelated point-wise displacements 
are learned. 

}

\vspace{-5pt}

\paragraph{\bf{Future work.}}
In addition to addressing the limitations discussed so far, 
a network capable of transforming point sets {\em hierarchically\/} is likely to produce more fine-grained results and adapt to more domains. 
We would also like to consider {\em transitive\/} transformations, where a source shape reaches a target via a sequence of two networks
through an intermediate shape. On this shape, the points can be upsampled, consolidated, filtered or undergo any other 
processing operation. This may also be generalized to combining and composing transformations. Finally, an intriguing avenue for future research
would be to relax the need for paired shapes, and replace it with an unsupervised or weakly supervised setting to train a general-purpose
network for point set transforms.

\section*{Acknowledgment}
The authors would like to thank the anonymous reviewers for their valuable comments. This work was supported in part by NSFC (61522213, 61761146002, 6171101466), 973 Program (2015CB352501), Guangdong Science Program (2015A030312015), Shenzhen Innovation Program (KQJSCX20170727101233642, JCYJ20151015151249564), ISF-NSFC Joint Research Program (2217/15, 2472/17), Israel Science Foundation (2366/16) and NSERC (611370).

\bibliographystyle{ACM-Reference-Format}
\bibliography{points}

\appendix
\section{Appendix}
\label{sec:appendix}

\subsection{Details of network architecture}
\label{sec:append_detail}

In this appendix, we provide details of the set abstraction layers, 
feature propagation layers, and fully connected layers in P2P-NET.
We use the same notations as in ~\cite{qi_nips2017}. 
A set abstraction layer of PointNet++ is denoted as $SA(K,r,[l_1,..,l_d])$, where $K$ 
is number of local patches, $r$ is radius of balls that bound the patches, $[l_1,..,l_d]$ 
are widths of fully connected layers used in local PoinNet. A feature propagation layer 
is denoted as $FP([l_1,..,l_d])$, where $[l_1,..,l_d]$ are widths of fully connected layers 
used inside the layer.  A fully connected layer is denoted as $FC(l)$, where $l$ is its 
width. Note that we disabled dropout for the FC layers.

For all experiments shown, we used the same A-P layers:
\begin{align*}
&\text{input} \rightarrow  SA( 1024, 0.1, [64, 64, 128] )  \rightarrow  SA( 384, 0.2, [128, 128, 256] )   \\
&\rightarrow  SA( 128, 0.4, [256, 256, 512] ) \rightarrow  SA( 1, 1.0, [512, 512, 1024] )  \\
&\rightarrow FP([512, 512])  \rightarrow FP([512, 256])  \rightarrow FP([256, 128]) \\
&\rightarrow FP([128, 128, 128]) \rightarrow \text{feature}
\end{align*}

We also used the same fully connected layers:
$$[\text{feature, noise}] \rightarrow  FC(128) \rightarrow FC(64)  \rightarrow  FC(3) \rightarrow \text{displacements}$$

\subsection{Closest training examples}
\label{sec:append_closetrain}

\begin{figure}[h]
\centering
\includegraphics[width=\linewidth]{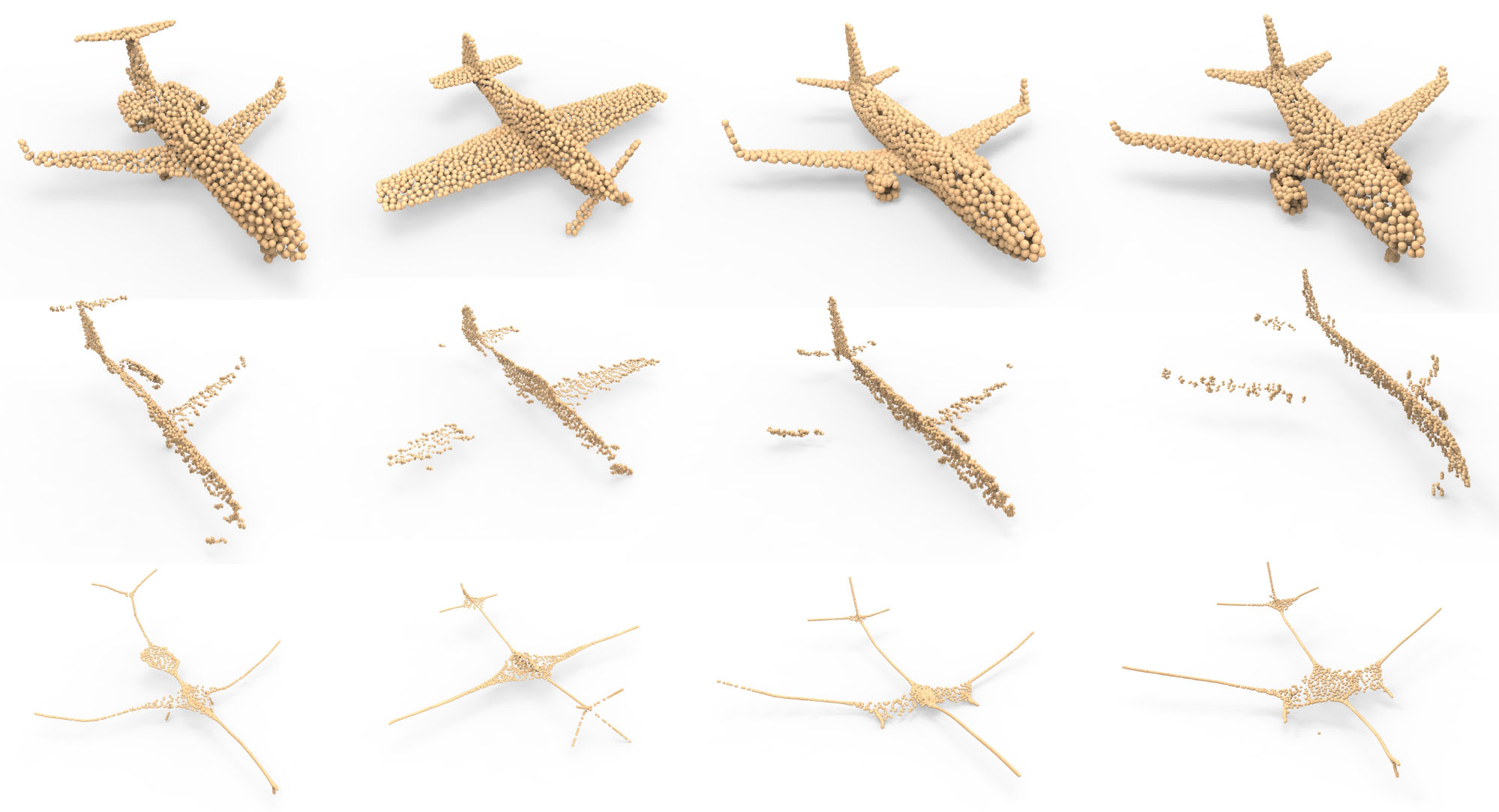}
\caption{Closest training examples for the test airplanes in Fig.~\ref{fig:SSStransform}.}
\label{fig:closest_airplane}
\end{figure}

\begin{figure}[h]
\includegraphics[width=\linewidth]{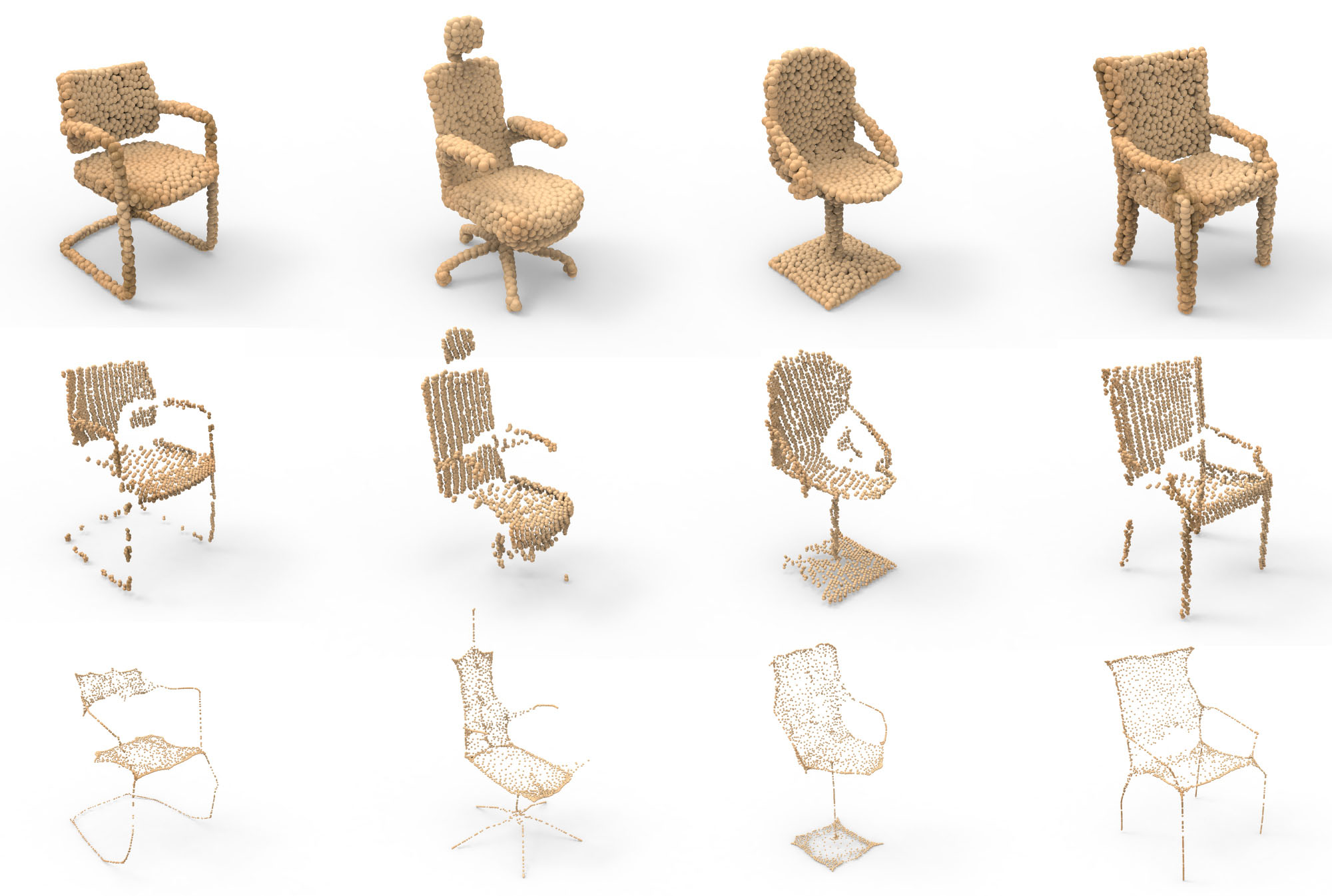}
\caption{Closest training examples for the test chairs in Fig.~\ref{fig:SSStransform}.}
\label{fig:closest_chair}
\end{figure}

In Figs.~\ref{fig:closest_airplane} and~\ref{fig:closest_chair}, we show the closest models from the training set
that are retrieved for the test examples used in Section~\ref{sec:skeletonshape}. The retrieval was done by 
searching for a training example having the closest surface samples to an input test example. To measure 
the difference between surface samples, we use the distance measure $D_\text{retrieve}(P,Q)$ as described 
in Section~\ref{sec:crossection}. 

\end{document}